%% file: main.tex
\documentclass[10pt,conference]{IEEEtran}

\usepackage{amsmath,amssymb,amsfonts}
\usepackage[hyphens]{url}
\usepackage{graphicx}
\usepackage{textcomp}
\usepackage[dvipsnames]{xcolor}
\usepackage{longtable}
\usepackage{todonotes}
\usepackage{multirow}
\usepackage{multicol}
\usepackage[many]{tcolorbox}
\usepackage{colortbl}
\usepackage{float}
\usepackage{placeins}
\usepackage[table,xcdraw]{xcolor}
\usepackage{enumitem}
\usepackage{soul}
\usepackage{breakurl}
\usepackage{balance}
\usepackage{algorithm}
\usepackage{algpseudocode}
\usepackage{breakurl}
\usepackage[breaklinks]{hyperref}

\tcbset{
    sharp corners,
    colback = white,
}
\renewcommand\st[1]{}

\newtcolorbox{boxD}{
    colback = {white},
    colframe = {black},
    boxrule = 0pt,
    toprule = 1.5pt,
    bottomrule = 1.5pt,
    left=3pt,right=3pt, top=2pt,bottom=2pt
}

\begin{document}

\title{A Story About Cohesion and Separation: Label-Free Metric for Log Parser Evaluation} %\heng{I feel label-free is better than unsupervised (for describing  model training process), but minor thing}

\author{
    \IEEEauthorblockN{
        Qiaolin Qin\IEEEauthorrefmark{1},
        Jianchen Zhao\IEEEauthorrefmark{2},
        Heng Li\IEEEauthorrefmark{1},
        Weiyi Shang\IEEEauthorrefmark{2},
        Ettore Merlo\IEEEauthorrefmark{1}
    }
    \IEEEauthorblockA{
        \IEEEauthorrefmark{1}Polytechnique Montreal, Montreal, Canada, \IEEEauthorrefmark{2}University of Waterloo, Waterloo, Canada
    }
    \IEEEauthorblockA{
    \{qiaolin.qin, heng.li, ettore.merlo\}@polymtl.ca, \{jianchen.zhao, wshang\}@uwaterloo.ca
    }

}

\maketitle
\input{sections/1_abstract}

\input{sections/2_introduction}
\input{sections/3_motivation}
\input{sections/4_methodology}
\input{sections/5_study_design}

\input{sections/6_experiments}
\input{sections/7_discussions}
\input{sections/8_related_works}
\input{sections/9_validity_threat}
\input{sections/10_conclusion}

\balance
\bibliographystyle{IEEEtran}
\bibliography{references}

\end{document}

%% file: sections/1_abstract.tex
\begin{abstract}
Log parsing converts log messages into structured event templates, allowing for automated log analysis and reducing manual inspection effort. To select the most compatible parser for a specific system, multiple evaluation metrics are commonly used for performance comparisons. However, existing evaluation metrics heavily rely on labeled log data, which limits prior studies to a fixed set of datasets and hinders parser evaluations and selections in the industry. Further, we discovered that different versions of ground-truth used in existing studies can lead to inconsistent performance conclusions. Motivated by these challenges, we propose a novel label-free template-level metric, PMSS (parser medoid silhouette score), to evaluate log parser performance. PMSS evaluates both parser grouping and template quality with medoid silhouette analysis and Levenshtein distance within a near-linear time complexity in general. To understand its relationship with label-based template-level metrics, FGA (F1-score of grouping accuracy) and FTA (F1-score of template accuracy), we compared their evaluation outcomes for six log parsers on the standard corrected Loghub 2.0 dataset. %According to the results, log parsers with the highest PMSS have 2.1\% average relative difference with the optimal FGA, and 9.8\% with optimal FTA. 
Our results indicate that log parsers achieving the highest PMSS or FGA exhibit comparable performance, differing by only 2.1\% on average in terms of the FGA score; the difference is 9.8\% for FTA.
PMSS is also significantly ($p<1e^{-8}$) and positively correlated to both FGA and FTA: the Spearman's $\rho$ correlation coefficient of PMSS-FGA and PMSS-FTA are respectively 0.648 and 0.587, close to the coefficient between FGA and FTA (0.670). We further extended our discussion on how to interpret the conclusions from different metrics, identifying challenges in using PMSS, and provided guidelines on conducting parser selections with our metric. PMSS provides a valuable evaluation alternative when ground-truths are inconsistent or labels are unavailable. 
\end{abstract} %\todo{specify what correlation}\heng{$O()$ is the algorithmic proof of the worst-case complexity. There should not be conclusions like ``in general or in most cases the complexity is $O(n)$''. Use ``linear'' if algorithmic proof is not possible}\heng{consider ``label-free'' vs. ``label-based''}

%% file: sections/2_introduction.tex
\section{Introduction}
\label{sec:introduction}

Log messages carry rich runtime information that is automatically recorded during system execution. This information is typically useful for understanding system behaviors and maintaining software quality. However, given the high volume of software log messages, it is infeasible to manually inspect each line of the log. Log parsing is a process of grouping similar log messages into event groups and extracting templates that preserve the semantic and structural features of corresponding events to automate downstream analysis~\cite{he2021survey}, such as log-based anomaly detection~\cite{kuttal2011history, gadler2017mining, fu2009execution, he2016experience, du2017deeplog, chen2021experience, wu2023effectiveness}, root cause analysis~\cite{lu2017log, yuan2012characterizing, fu2013contextual}, or performance regression detection~\cite{liao2020using, chow2014mystery, nagaraj2012structured}. 

Many log parsers have been proposed to perform automatic log parsing over the years, and evaluation metrics are required to understand and compare their parsing effectiveness. Mainstream log parser effectiveness metrics include grouping accuracy (GA)~\cite{zhu2019tools}, parsing accuracy (PA)~\cite{dai2020logram}, F1-score of grouping accuracy (FGA)~\cite{jiang2024large}, and F1-score of template accuracy (FTA)~\cite{khan2022guidelines}. GA and PA evaluate the correctness of log grouping and template abstraction on the message level (i.e., the ratio of messages correctly grouped/abstracted into a template), while FGA and FTA evaluate a parser's effectiveness on the template level (i.e., the F1-scores of templates correctly grouped/abstracted). Their justification of correctness is always based on labeled ground-truths. %In other words, labeled log data is always needed to evaluate log parsers. %\heng{Briefly explain these metrics since they are used extensively before their definitions in 4.3} 

Although these label-based metrics can indeed reveal the performance of log parsers, we found that their reliance on ``ground-truth templates'' brought three challenges in evaluation. First, the \textit{real} ground-truth event templates encoded as logging statements are often inaccessible in practice~\cite{zhu2023loghub,he2021survey,jiang2024large}. The manual, labor-intensive template labeling process limits the number of available benchmark datasets in academic research and industrial implementation. Second, we noticed that researchers hold diverse opinions on template correctness, which leads to diverse template correction rules that create different ground-truth versions~\cite{khan2022guidelines,xiao2024free,huang2025no,le2025unleashing}. Nevertheless, label ``correctness'' is often unverifiable without logging statements. Third, a previous study by Khan et al.~\cite{khan2022guidelines} revealed that different ground-truth versions can lead to inconsistent conclusions on parser performances. Our experimental results also confirmed this finding: the evaluation scores obtained from different ground-truth versions vary largely, and the best-performing parser may not hold across versions. 

These challenges in label-based parser evaluation motivate us to seek label-free parser effectiveness metrics. By definition~\cite{zhu2023loghub,he2021survey,jiang2024large}, log parsing shares common natures with two widely studied tasks in natural language processing, namely text clustering (for log grouping) and text summarization (for template quality evaluation). Both tasks can be evaluated with label-free approaches.% Further, since unsupervised K-means text clustering evaluations require distance metrics, which are often defined with text similarity/distance, we argue that \textbf{log parser performance can be assessed with a single clustering metric}. \heng{I also could not understand the logic here. Something is missing. }

In this study, we propose the parser medoid silhouette score (PMSS) to jointly evaluate parser grouping and template quality. Medoids are robust representations of cluster centers~\cite{van2003new}; PMSS defines the templates as medoids (i.e., a candidate actual logging statement) to evaluate the average cohesion (i.e., the distance between each log message and its corresponding template) and separation (i.e., the distance between each log message and its neighbor templates) among event groups. The distances between messages and templates are defined with the Levenshtein distance~\cite{lcvenshtcin1966binary} to capture their semantic-structural difference. A high PMSS value indicates that the parser can generate templates describing their matched messages while separating different event groups. Through experiments, we discuss the relationship between PMSS and the mainstream label-based template-level metrics (i.e., FGA and FTA). Then, we provide guidelines on how to use PMSS to comprehend log parser performances and select the most suitable parser for downstream tasks. We answer the following three research questions in our study:
%\heng{It reads confusing. If EMSS is not for the main results/conclusions, then only mention it in discussion to avoid disturbing the main storyline. The term} We also provide an event-level\heng{Only until Algorithm 1 I understood what event-level means in the paper. Since an ``event'' is the same thing as a ``template'' in most log analysis papers, eevnt-level (including the EMSS metric mentioned below, could be named as individual-template-level to avoid confusion.} parsing quality metric, event medoid silhouette score (EMSS), for fine-grained parser performance analysis. 

\noindent\textbf{RQ1: How do inconsistencies among log data ground-truth versions influence the reliability of parser evaluation results?} We collect four sets of different template correction rules from recent studies~\cite{jiang2024large,huang2025no,xiao2024free,le2025unleashing} and examine how they influence the consistency of log parsing evaluations. We noticed that a tool's performance can be vastly different on the same dataset with varied ground-truth versions: on the OpenStack dataset, LUNAR's min-max PA score difference is 0.517, where the maximum is 216.2\% of the minimum. The score shifts among ground-truth versions also lead to divergent conclusions on the most compatible tool for a certain dataset, which hinders the parser selection process in practice. 
%\heng{The details from here until before ``We noticed that a tool's performance can be vastly different'' can be removed. We only need to provide a high-level introduction of the RQ here}These functions are applied to the original version of Loghub 2.0~\cite{jiang2024large}, the dataset adopted in most of these studies (LogBatcher used its smaller variant, Loghub-2k), to obtain different ground-truth versions. We observed that the implementations of correction rules vary from each other, and LUNAR's ground-truth has the largest difference at both message and template levels with the other versions. Further, we obtained one-run parsing results for six state-of-the-art log parsers on Loghub 2.0 and evaluated them with mainstream label-based metrics (i.e., GA, PA, FGA, and FTA) on the five ground-truth versions.

\noindent\textbf{RQ2: What is the relationship between PMSS and FGA-FTA?} To validate the effectiveness of PMSS, we compare its evaluation results for the studied log parsers and compare them with those obtained from using label-based template-level metrics. The best parser selected by the PMSS metric is also the best one selected by FGA in 7 and 14 datasets. The Spearman's $\rho$ analysis also suggests that PMSS is significantly and positively related to both FGA and FTA. Nevertheless, since PMSS evaluates grouping and parsing performance with template-message similarity instead of templates' identicalness with ground-truth, it does not guarantee the highest score for a tool with both optimal FGA and FTA. 
%We calculated the PMSS scores for the evaluated tools using the same results obtained in RQ1 and discussed their relationships with 
 % calculated on the Loghub 2.0 ground-truth corrected with the rules provided in its original repository~\cite{jiang2024large}. 
%The tools with optimal PMSS have an average relative difference of 2.1\%\heng{this is not clear in the abstract and here: what is the relative difference? Why not use a more straightforward way, like: the best parser selected by the PMSS metric is also the best one selected by FGA and FTA for X and Y out of Z datasets, respectively, or, the best  parser selected by the PMSS metric is also the best one selected by either FGA or FTA for X out of Z datasets.} from optimal FGA, and 9.8\% in terms of FTA. 

\noindent\textbf{RQ3: How efficient is the calculation of PMSS in comparison to FGA and FTA?} We compared the evaluation time of our metrics to the two template-level label-based metrics (i.e., FGA and FTA) to understand their efficiency. Without the aid of ground-truths, PMSS calculation requires additional data processing, such as regex matching, and is generally slower than FTA and FGA calculations on most of the datasets. Nonetheless, PMSS retains a linear time complexity in the majority of cases, which ensures a reasonable and stable evaluation time. Its average evaluation time per 1,000 lines is generally around 15ms on our studied datasets.

%ranges from 0.38 to 287.07 seconds on our studied datasets. %\heng{could mention the range of execution time for the studied datasets}

Our work makes the following main contributions: 

\begin{enumerate}
    \item We systematically examined the difference among template correction rules and how they influence log parser evaluation. The findings revealed challenges in using label-based metrics for parser comparison. 
    \item We proposed parser medoid silhouette score (PMSS), a label-free template-level effectiveness metric. By combining unsupervised clustering analysis and text similarity calculation, the metric provides a comprehensive assessment of both parser grouping and template quality. %\heng{the details below are not necessary (already mentioned above)}The metric significantly correlates with both supervised template-level metrics, FGA and FTA. Specifically, we found that tools with the highest PMSS have a small average relative difference (2.1\%) with the optimal tool in terms of FGA. Further, we propose the event medoid silhouette score (EMSS) to provide finer-grained insights into the parsing quality of each event group. 
    \item We also compared the efficiency of PMSS with FGA and FTA. Although PMSS requires more computational time to process the parsing results, it still maintains a linear complexity in the majority of cases, ensuring efficient and stable parser evaluation. 
\end{enumerate}

The experimental code and data are publicly available for replication~\cite{PMSS}. 

%\noindent\textbf{Paper Organization.} %
The remainder of our paper is organized as follows. 
In Section~\ref{sec:motivation}, we discuss the three challenges in label-based log parser evaluation and existing approaches for unsupervised text clustering analysis. 
In Section~\ref{sec:approach}, we introduce the calculation process of PMSS. 
In Section~\ref{sec:experiment_settings}, we introduce the implementation details of our experiments. 
In Section~\ref{sec:results}, we present our experimental results to answer the research questions. 
In Section~\ref{sec:discussions}, we extend the discussions on PMSS's relationship with mainstream label-based metrics and provide guidelines for applying PMSS for parser selection. 
In Section~\ref{sec:related_works}, we introduce existing studies related to our work, and in Section~\ref{sec:validity_threat}, we discuss the threats to the validity and our mitigation strategies. Finally, we summarize the findings and contributions of our work and propose future work in Section~\ref{sec:conclusion}. %contribution provided by our research.  

%% file: sections/3_motivation.tex
\section{Motivation}
\label{sec:motivation}

\subsection{Challenges in Label-based Log Parser Evaluation}
\label{sec:challenges}

\noindent\underline{\textbf{Challenge 1: Limited Availability of Ground-truth.}}

The \textit{real} ground-truth event templates for log messages are determined by their logging statements ~\cite{zhu2023loghub,he2021survey}, which are often inaccessible due to large codebases or cross-team access restrictions~\cite{jiang2024large, khan2024impact}: a recent study by Khan et al.~\cite{khan2024impact} found that only 3 systems' (HDFS, Hadoop, OpenStack) logging statements are available among the 16 systems involved in Loghub-2k~\cite{zhu2023loghub}. As an alternative, researchers manually label event templates for a large number of log messages to create datasets for parser evaluation. The labor-intensive log labeling process \textbf{limits related research} to a fixed set of datasets. 

The labeling requirement also restricts parser evaluation in industrial scenarios. Since ground-truth templates are often unavailable for large-scale commercial systems, log parsers are often compared with average performance on open-source benchmarks. However, an average optimum does not promise optimal compatibility with a specific dataset. For instance, when evaluated with the officially corrected Loghub 2.0 ground-truth~\cite{jiang2024large}, LUNAR obtained a higher average FTA (0.736) than LogBatcher (0.718), while LogBatcher's FTA on the HDFS dataset is 0.946, largely outperforming LUNAR (0.731). The metrics' reliance on labeled ground-truths also \textbf{hinders the selection of suitable log parsers in industrial scenarios}. The constraints imposed on both research and industrial implementations call for label-free evaluation metrics. 

\noindent\underline{\textbf{Challenge 2: Mislabeling and Diverged Correction Rules.}}

Although the datasets commonly used in existing research are labeled by a group of domain experts, the manual process may inevitably introduce errors~\cite{jiang2024large,khan2022guidelines,le2025unleashing}. Khan et al.~\cite{khan2022guidelines} argued that the templates in Loghub-2k~\cite{zhu2023loghub} contain issues such as ``fixed parts that are unlikely to be hard-coded in log printing statements''. To improve template quality, they stressed the importance of template correction and proposed 8 heuristic correction rules (e.g., merging consecutive placeholders), which are later applied in the Loghub 2.0 repository~\cite{jiang2024large}.

However, comprehension differences can lead to diverse template correctness definitions and varied correction rules. Xiao et al. argue that the heuristic rules proposed by Khan et al.~\cite{khan2022guidelines} are not comprehensive. Thus, they added several dataset-specific rules (e.g., standardizing ``tty=ssh'' and tty=NodeVssh'' to ``tty=\textless*\textgreater'' in Linux and OpenSSH) for Loghub-2k correction~\cite{xiao2024free}. Instead of leveraging the standard correction rules in Loghub 2.0~\cite{jiang2024large}, Le et al. practiced several light-weight corrections to evaluate UNLEASH~\cite{le2025unleashing}. On the other hand, Huang et al. applied additional heuristics based on the Loghub 2.0's standard heuristics to evaluate LUNAR~\cite{huang2025no}.  

%While the standard version of Loghub 2.0~\cite{jiang2024large} adopted the 8 heuristic rules proposed by Khan et al.~\cite{khan2022guidelines} to correct the raw labeled files, 

The misalignment in correction rules shows that researchers \textbf{lack consensus on event template correctness}. Since logging statements are generally not available, \textbf{the correctness of templates cannot often be verified}. The ground-truth correctness issues for log parsing further motivate us to propose a label-free evaluation technique. 

\noindent\underline{\textbf{Challenge 3: Potential Conclusion Inconsistency.}}

\input{tables/template_merge}

Khan et al.~\cite{khan2022guidelines} showed that subtle score shifts in different ground-truth versions can lead to varied conclusions in parser performance rankings. For example, on Loghub-2k, Drain~\cite{he2017drain} gained 0.29 in average parsing accuracy (PA) using the original labels, and 0.34 after template corrections, surpassing SLCT~\cite{vaarandi2003data} and becoming the overall most effective log parser among the compared ones. Apart from their impact on template quality evaluations, these corrections may also lead to event merging that change the grouping labels and evaluation results. As illustrated by the example in Table~\ref{tab:template_merge_example}, after applying the value assignment rule, the two previously separated events $E_1$ and $E_2$ will merged into $E'$. 

Ground-truths are the ``gold standard'' reference for tool performance comparison. If all ground-truths are accurate and complete, the conclusions on both perspectives should always be identical across versions, even if absolute metric values vary slightly due to minor annotation differences. Nevertheless, based on the discoveries of Khan et al., \textbf{inconsistencies in conclusions may occur} when evaluations are conducted with different unverified ground-truth versions. This concern motivates us to evaluate the performance discrepancies for each tool across ground-truth versions, as well as the parser performance ranking changes on the same dataset. 

\subsection{Unsupervised Text Clustering Evaluation}
\label{sec:unsupervised_clustering_analysis}

Log parsing is a process of \textbf{grouping log messages with the same event semantics and abstracting event templates that contain key event information}~\cite{he2021survey,jiang2024large,khan2022guidelines,zhu2023loghub}. By definition, log parsing integrates two analogous tasks from natural language processing: log grouping is similar to a text clustering process; template extraction from log groups resembles an unsupervised text summarization task, where the goal is to derive a concise structural abstraction that represents the common semantics of the grouped messages. Given their similar nature, we can conduct label-free log parser evaluations with metrics adopted from these fields. 

%as an unsupervised K-means-like\heng{use ``partitioning clustering such as K-means'' to be more general} 

Unsupervised clustering evaluations rely on a predefined distance or similarity function. In text clustering, this function is typically defined using text similarity measures, such as Jaccard distance or cosine similarity over TF-IDF representations. Similarly, unsupervised text summarization evaluation relies on text similarity metrics to quantify the semantic alignment or content overlap between the source texts and their generated summaries. Hence, we propose a metric that combines the two evaluations based on their shared distance calculation. 
%Based on this reasoning, \textbf{both log parser grouping and template quality can be jointly evaluated within an unsupervised clustering metric, where distances are computed using text similarity metrics}.\heng{The reasoning here is not convincing: they share the distance metric does not directly mean they can be evaluated together. It's better to just say we propose a metric that combine the two metrics based on their shared distance calculation.}

Without ground-truth labels, clustering quality is often assessed in two ways. The first type of approach (e.g., silhouette analysis) examines each data point’s cohesion (i.e., distance to its assigned cluster) and separation (i.e., distance to the nearest other cluster). A desirable clustering should ensure that each data point is closer to its assigned cluster than to any other cluster. The second type of approach (e.g., the elbow method) evaluates the trade-off between the number of clusters (i.e., model complexity) and the average in-cluster distance, aiming to balance between model simplicity and cluster compactness. However, their identification of optimal balance is often subjective, and they cannot justify the cluster separation performance. Given that the fundamental requirement for log parsing is to aggregate similar log messages while separating different events~\cite{he2021survey}, the evaluation of log parsing is more in line with the first approach. 

A recent survey by Wang and Dong investigated statistic-based and semantic-based text similarity calculation approaches in the field of natural language processing~\cite{wang2020measurement}. The semantic-based approaches leverage language models to encode texts and calculate their vectors' similarity, while the statistic-based ones compute similarity based on word coexistence counts or probabilities. In comparison, statistic-based approaches are more lightweight and thus are more suitable for evaluating large volumes of log parsing data. The extracted event templates should preserve both the semantic and structural attributes of their corresponding log messages. Specifically, the tokens in a template should appear in the original messages in an identical sequence. To quantify this alignment, the Levenshtein distance~\cite{lcvenshtcin1966binary} provides a suitable measure of template–message distance. 

Building on the above reasoning, we propose a label-free log parsing evaluation metric that combines the medoid-based silhouette score~\cite{van2003new} with the Levenshtein distance to jointly assess log parser's grouping and template extraction quality.

%% file: tables/template_merge.tex
% \usepackage{graphicx}
\begin{table}[]
\caption{An example of event template merging after applying the value assignments (VA) correction rule. }
\label{tab:template_merge_example}
\resizebox{\columnwidth}{!}{%
\begin{tabular}{l|l}
\hline
\textbf{$E_1$} & logname= uid=\textless$*$\textgreater euid=\textless$*$\textgreater tty=\textcolor{red}{NODEVssh} ruser= rhost=\textless$*$\textgreater user=\textcolor{red}{test}    \\ \hline
\textbf{$E_2$} & logname= uid=\textless$*$\textgreater euid=\textless$*$\textgreater tty=\textcolor{red}{NODEVssh} ruser= rhost=\textless$*$\textgreater user=\textcolor{red}{guest}   \\ \hline
\textbf{$E'$} & logname=\textcolor{red}{\textless$*$\textgreater} uid=\textless$*$\textgreater euid=\textless$*$\textgreater tty=\textcolor{red}{\textless$*$\textgreater} ruser=\textcolor{red}{\textless$*$\textgreater} rhost=\textless$*$\textgreater user=\textcolor{red}{\textless$*$\textgreater} \\ \hline
\end{tabular}%
}
\end{table}

%% file: sections/4_methodology.tex
\section{Parser Medoid Silhouette Score}
\label{sec:approach}

PMSS (parser medoid silhouette score) is the average value of EMSS (event medoid silhouette score), which evaluates the message-template coherence per event.  The calculation of PMSS comprises five steps. We present the pseudocode for our approach in Algorithm~\ref{alg:pseudocode}, and we elaborate on each step in the following subsections. 

\input{tables/pseudo_code}
\input{tables/tokenization_example}

\subsection*{\emph{\textbf{Step 1: Template Preparation}}}
\label{sec:template_preparation}
The medoid silhouette score is calculated with the distances between the data points' assigned cluster center and their neighboring cluster center (i.e., the closest cluster where the examined data point is not a member). Following this strategy, the time complexity will be $O(m*n)$, where $n$ denotes the number of log messages and $m$ denotes the number of templates. In the worst case, the complexity can be $O(n^2)$ when each template is identical to the original message. 

We therefore sort the event templates alphabetically to reduce time complexity. The intuition is that if a message shares high semantical and structural similarity with its log template, then it would also share more similarity with the sorted neighbors in comparison to ``farther'' templates. Then, we remove the pure numerical elements (\textbackslash b(\textbackslash d+)\textbackslash b) in the templates according to the heuristic of He et al., that these elements are always variables~\cite{he2017drain}. To facilitate further log message processing, we convert the template $t_{E_i}$ for each group $E_i$ into a regex $rex_{E_i}$: the template example in Table~\ref{tab:tokenization_example} will be converted to ``Bluetooth\textbackslash: (.*?) \textbackslash(ver (.*?)\textbackslash)''. Afterward, we tokenize $t_{E_i}$ into a token list $lst_{E_i}$ with placeholders (``\textless$*$\textgreater'') and blank spaces, as shown in Table~\ref{tab:tokenization_example}. 

\subsection*{\emph{\textbf{Step 2: Message Matchability Verification}}}
\label{sec:message_matchability_verification}

Given that log messages are automatically generated with logging codes, they must be matchable (e.g., using regex) by their corresponding templates. For a log message $log_i$ grouped into an event group $E_j$, we test matchability $Match(log_i,rex_{E_j})$ by verifying whether the message matches its template's corresponding regex $rex_{E_j}$. 

\subsection*{\emph{\textbf{Step 3: Message Tokenization}}}
\label{sec:message_tokenization}

The message will then be tokenized as $lst_{log_i}$ if it matches $rex_{E_j}$ in Step 2. The constant parts of the template (i.e., the strings segmented by placeholders) are used to locate dynamic parts in the log message. The separation of constant and dynamic parts in log messages can prevent the distance calculation bias introduced by constant tokens mixed with variables (e.g., users=\textless$*$\textgreater). After separating the two parts, we split the strings with blank spaces and ordered the tokens according to their sequences in the original message. We provide a message tokenization example in Table~\ref{tab:tokenization_example}. 

\subsection*{\emph{\textbf{Step 4: Message Silhouette Coefficient Calculation}}}
\label{sec:message_sc_calculation}

After tokenization, we use the Levenshtein distance (i.e., edit distance with insertion, deletion, and substitution)~\cite{lcvenshtcin1966binary} to calculate the message's distance to its corresponding template (inner distance, $Dist_{in}(log_i)$) and its neighbor templates. The outer distance $Dist_{out}(log_i)$ is defined as the smallest Levenshtein distance between the message and its two neighbor templates. The silhouette coefficient for each matched log message (i.e., $Match(log_i,rex_{E_j})=True$) is calculated as:
\[
s(log_i)=\frac{Dist_{out}(log_i)-Dist_{in}(log_i)}{max(Dist_{out}(log_i), Dist_{in}(log_i))}
\]
The log messages whose $Match(log_i,rex_{E_j})=False$ do not share structural similarity with their corresponding template. Hence, we set their silhouette coefficients $s(log_i)=0$, indicating that the template cannot better represent (i.e., have a small distance to) the message than its neighbor templates. 
 
\subsection*{\emph{\textbf{Step 5: Silhouette Score Calculation}}}
\label{sec:parser_ss_calculation}

We propose PMSS (parser medoid silhouette score) and EMSS (event medoid silhouette score) to understand the parser's performance on different levels. The two scores are calculated as follows:
\[
EMSS(E_j)=\frac{\sum_{log_i\in E_j}s(log_i)}{|E_j|},\  j=1,...,m
\]
\[
PMSS=\frac{\sum_{E_i}^mEMSS(E_i)}{m}
\]
where $|E_j|$ stands for the number of log messages in the event group $E_j$. EMSS provides a fine-grained understanding of whether each event template is highly coherent with the corresponding messages in terms of semantics and structure. PMSS is a template-level metric that provides an overview of the parser's performance across all events. A higher PMSS value indicates that the parser can generally obtain better in-event cohesion and inter-event separation. 

%% file: tables/pseudo_code.tex
\newcommand{\StepBlock}[1]{\Statex \textbf{#1}}
\begin{algorithm}
    \caption{Parser Medoid Silhouette Score Calculation}
    \label{alg:pseudocode}
    \begin{algorithmic}
    \item[\textbf{Input:}] $logs=\{log_1,...,log_n\}$, \Statex \hspace{2em}$event\ templates=\{t_{E_1},...,t_{E_m}\}$
    \item[\textbf{Output:}] $\{EMSS(E_1),...,EMSS(E_m)\}$, $PMSS$ 
    %\State $rex\gets [\ ]$, $lst_{temps} \gets[\ ]$
    \StepBlock{/*  Step 1: Template Preparation  */}
    \State Sort $\{t_{E_1},t_{E_2},...,t_{E_m}\}$ alphanumerically.
    \For{$t_{E_i}\in temps$}
    \State $rex_{E_i}\gets Regex(t_{E_i})$, $lst_{E_i}\gets Tokenize(t_{E_i})$
    %\State $lst_{temps}.append(lst_{E_i})$, $rex.append(rex_{E_i})$
    \EndFor
    \StepBlock{/*  Step 2-4: Log Message Processing  */}
    \For{$log_i\in logs$, $log_i$ assigned to $E_j$}
    \If{$Match(log_i,rex_{E_j})$} 
        \State $lst_{log_i}\gets Tokenize(log_i,lst_{E_j})$
        \Statex\hspace{2.8em} // \textit{$LDist(x,y)$ calculates Levenshtein distance}
        \State $Dist_{in}(log_i)\gets LDist(lst_{log_i},lst_{E_j})$ 
        \State $Dist_{out}(log_i)\gets min(LDist(lst_{log_i},lst_{E_{j+1}})$, \Statex \hspace{9.5em} $LDist(lst_{log_i}, lst_{E_{j-1}}))$
        \State $s(log_i)\gets\frac{Dist_{out}(log_i)-Dist_{in}(log_i)}{max(Dist_{out}(log_i),Dist_{in}(log_i))}$
    \Else
        \State $s(log_i)=0$
    \EndIf     
    \EndFor
    \StepBlock{/*  Step 5: Silhouette Score Calculation  */}
    \For{$E_j\in events$}
    \State $EMSS(E_j)\gets\frac{\sum_{log_i\in E_j }s(log_i)}{|E_j|}$
    \EndFor
    \State $PMSS\gets\frac{\sum_{E_i}^mEMSS(E_i)}{m }$
    \end{algorithmic}
    
\end{algorithm}

%% file: tables/tokenization_example.tex
% \usepackage{graphicx}
\begin{table}[]
\caption{An example for template and message tokenization. }
\label{tab:tokenization_example}
\resizebox{\columnwidth}{!}{%
\begin{tabular}{l|l|l}
\hline
               & \multicolumn{1}{c|}{\textbf{Original}} & \multicolumn{1}{c}{\textbf{Tokenized}}      \\ \hline
\textbf{Temp.} & Bluetooth: \textless$*$\textgreater  (ver \textless$*$\textgreater)           & ['Bluetooth:', '(ver', ')']                 \\
\textbf{Msg.}  & Bluetooth: L2CAP (ver 2.1)             & ['Bluetooth:', 'L2CAP', '(ver', '2.1', ')'] \\ \hline
\end{tabular}%
}
\end{table}

%% file: sections/5_study_design.tex
\section{Experiment Settings}
\label{sec:experiment_settings}

\subsection{Dataset}
\label{sec:dataset}
To study the impact of different template correction rules on label-based log parser evaluation and understand the relationships between label-based and label-free metrics, we use the Loghub 2.0~\cite{jiang2024large} dataset, an extended version of Loghub-2k~\cite{zhu2023loghub}, for our experiments. Loghub 2.0 contains high volumes of annotated log messages (23,921 to 16,601,745) and various templates (11 to 1,241). The 14 log datasets in Loghub 2.0 are originally sampled from different sources, including distributed systems (Hadoop, HDFS, OpenStack, Spark, and Zookeeper), supercomputer systems (BGL, HPC, and Thunderbird), operating systems (Linux and Max), server-side applications (Apache and OpenSSH), and standalone software (HealthApp and Proxifier). 

The original Loghub 2.0 data is downloaded directly from Zenodo~\cite{loghub2.0}. Then, we used the template correction functions from Loghub 2.0~\cite{zhu2023loghub}, LogBatcher~\cite{xiao2024free}, UNLEASH~\cite{le2025unleashing}, and LUNAR~\cite{huang2025no} to create five versions of ground-truths for impact evaluation (including the original, uncorrected version).

\subsection{Log Parser Selection}
\label{sec:parser_selection}
Our study focuses on investigating the effect of ground-truth version differences on evaluation outcomes. To this end, we selected six state-of-the-art open-source unsupervised log parsers for our experiments.

% \noindent\underline{\textbf{Drain}}
\noindent\underline{\textbf{Drain}}~\cite{he2017drain}. Drain is a representative state-of-the-art log parser, which demonstrated high effectiveness and efficiency in comparison to other statistics-based parsers~\cite{jiang2024large}. We use Drain's default code provided in Loghub 2.0 for implementation. 

\noindent\underline{\textbf{Preprocessed-Drain}}~\cite{qin2025preprocessing}. A recent study by Qin et al. further improved Drain's effectiveness through log preprocessing. We use their replication code with default preprocessing regexes. 
    
\noindent\underline{\textbf{LILAC}}~\cite{jiang2024lilac}. LILAC is a semantic-based parser that leverages GPT-3.5-Turbo-0613 for log parsing. We used a 0-shot prompting strategy with GPT-3.5-Turbo-0125 during our replication because GPT-3.5-Turbo-0613 is no longer available. 

\noindent\underline{\textbf{LibreLog}}~\cite{ma2024librelog}. LibreLog uses smaller-scale open-source LLMs for log parsing. Due to hardware limitations, we did not implement this tool in our experiment. Its parsing results on Loghub 2.0 using Llama3-8B (i.e., the best performance setting in the original study) are obtained from the authors. For a unified evaluation, we converted the template regexes in LibreLog to the same format as Loghub 2.0 event templates. 

\noindent\underline{\textbf{LogBatcher}}~\cite{xiao2024free}. LogBatcher further enhanced the effectiveness and efficiency of semantic-based log parsing.  We leveraged the code with the default settings (i.e., using GPT-3.5-Turbo-0125) from the paper to obtain the results. 

\noindent\underline{\textbf{LUNAR}}~\cite{huang2025no}. LUNAR leverages a new strategy based on log commonality and variability for grouping to achieve better parsing effectiveness. We implemented their code with default settings (i.e., using GPT-3.5-Turbo-0125) for parsing. 

We intentionally excluded supervised log parsers (e.g., few-shot LILAC \cite{jiang2024lilac} and UNLEASH \cite{le2025unleashing}) because they rely on few-shot prompting with samples drawn from the ground-truth, which introduces an additional source of randomness, since different sampled labels can influence model prediction. Including such tools would confound the effects of ground-truth version differences with sampling and model variability. By restricting our study to unsupervised parsers, whose outputs only rely on input logs, we ensure that any observed variation in evaluation results can be attributed solely to the choice of ground-truth version, which aligns with our research objective.

\subsection{Baseline Metrics}
\label{sec:metrics}
In addition to PMSS, we selected four mainstream effectiveness metrics in our experiments, namely GA, PA, FGA, and FTA. The metric scores are calculated with the standard codes provided in Loghub 2.0~\cite{jiang2024large}.

\noindent\underline{\textbf{Grouping Accuracy (GA)}}~\cite{zhu2019tools}. GA is a message-level metric defined as the ratio of correctly grouped log messages in a dataset. A log message is correctly grouped if it is clustered with all other messages sharing the same ground-truth event. 

\noindent\underline{\textbf{Parsing Accuracy (PA)}}~\cite{dai2020logram}. PA is a message-level metric defined as the ratio of correctly parsed log messages in a dataset. A log message is correctly parsed if its abstracted template is identical to the ground-truth template. 

\noindent\underline{\textbf{F1-score of Group Accuracy (FGA)}}~\cite{jiang2024large}. FGA is a template-level metric, defined with the harmonic mean of the grouping precision (PGA) and grouping recall (RGA). An event is correctly grouped if the group only contains all the log messages sharing the same ground-truth template. %\todo{the equations could be made inline if space is not enough} Since it does not consider the message frequency, FGA offers the same weight to both frequent (e.g., running status) and infrequent log templates (e.g., error reports). $N_c$ is the number of events correctly grouped by the parser, $N_p$ is the number of total events identified by the parser, and $N_g$ is the number of ground-truth events. $\hat{N_c}$ is the number of correctly identified templates. 

\noindent\underline{\textbf{F1-score of Template Accuracy (FTA)}}~\cite{khan2022guidelines}. FTA is a template-level metric, defined with the harmonic mean of the template precision PTA and template recall RTA. A template is correctly identified when the event is correctly grouped and the event's template is identical to the ground-truth. 

\subsection{Experiment Environment}
\label{sec:experiment_environment}
We obtained LibreLog's parsing results from its authors; thus, its experimental environment is aligned with the original paper: LibreLog is implemented with an Ubuntu server with a NVIDIA Tesla A100 GPU, AMD EPYC 7763 64-core CPU, and 256GB RAM~\cite{ma2024librelog}. We run the remaining five log parsers and evaluate the six parsing outcomes on a Mac Mini with an M2 chip and 16GB of memory.

\subsection{Experiment Process}
\label{sec:experiment_process}

\begin{figure}[!t]
  \center
  \includegraphics[keepaspectratio=1,width=\columnwidth]{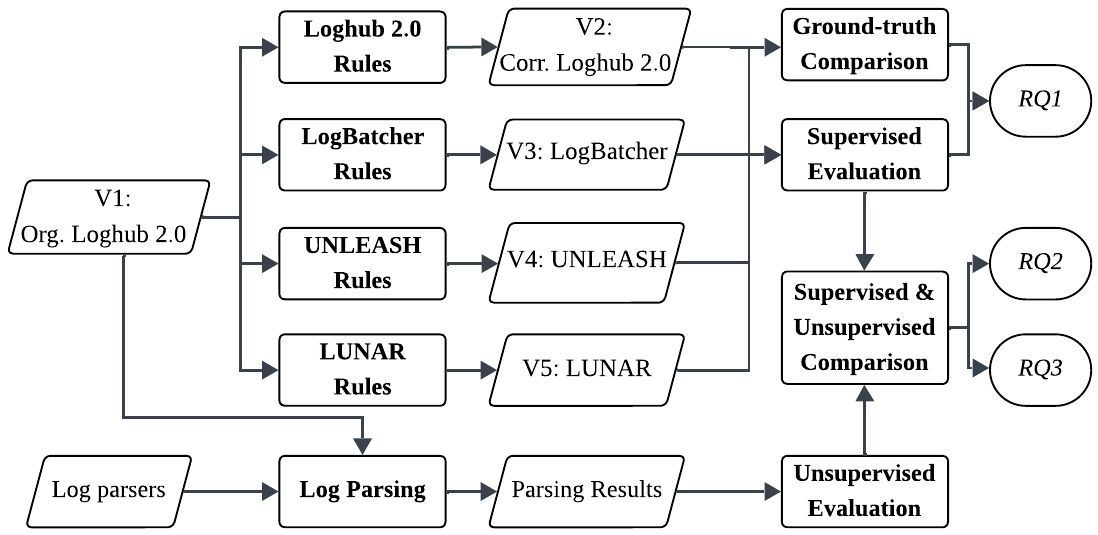}
  \caption{The experiment process overview.}
  \label{fig:overview}
\end{figure}

\input{tables/correction_rules}

Our study comprises five main steps to answer the three research questions proposed in Sec.~\ref{sec:introduction}. The overview of the process is shown in Figure~\ref{fig:overview}. We elaborate on the details of conducting each step below.

\noindent\textbf{1) Data Collection.} We collected four different template correction functions from the replication package of Loghub 2.0~\cite{jiang2024large}, LogBatcher~\cite{xiao2024free}, UNLEASH~\cite{le2025unleashing}, and LUNAR~\cite{huang2025no}. The four functions are independently implemented on the original Loghub 2.0 dataset to create four corrected ground-truths, namely original Loghub 2.0 (V1), corrected Loghub 2.0 (V2), LogBatcher (V3), UNLEASH (V4), and LUNAR (V5). 

\noindent\textbf{2) Log parsing.} After obtaining LibreLog's results on Loghub 2.0 from its authors, we converted the template regex results to Loghub-style event templates. Then, we used the remaining 5 log parsers to parse the Loghub 2.0 dataset and collected their parsing results from one run.

\noindent\textbf{3) Efficiency Evaluation.} First, we evaluate the parsing results with GA, PA, FGA, and FTA on the five ground-truth versions. Then, we leveraged our label-free metric, PMSS, to evaluate log parser effectiveness using the same parsing outputs. 

\noindent \textbf{4) Ground-truth and Label-based Metrics Comparison.} To answer RQ1, two authors independently categorized the correction rules according to the descriptions in their implementation codes. The final categories are based on the result of their discussion. Then, we calculated the ratio of events and messages having different templates between ground-truth version pairs. Afterward, we compared the label-based metric outcomes obtained from the five ground-truth versions. Following the experiment strategy of Khan et al.~\cite{khan2022guidelines}, we examine their differences from two aspects: metric score min-max discrepancies and optimal-performance parser changes. 

\noindent\textbf{5) Label-based and Label-free Metrics Comparison.} Since PMSS evaluate parser performance on the template level, we discuss its relationship with the two template-level label-based metrics, FGA and FTA. The FGA and FTA scores being compared are calculated using the standard corrected Loghub 2.0 ground-truth (V2). In RQ2, we compare their effectiveness evaluation outcome. In RQ3, we compare their evaluation efficiencies based on the calculation time of the metric.

%% file: tables/correction_rules.tex
% \usepackage{graphicx}
\setlength{\tabcolsep}{2pt}
\begin{table*}[]
\centering
\caption{The 10 template correction rules collected from the four correction functions and their implementations in different ground-truth versions.}
\label{tab:correction_rules}
\resizebox{\textwidth}{!}{%
\begin{tabular}{l|l|l|c|c|c|c}
\hline
\textbf{Rule} &
  \textbf{Description} &
  \textbf{Example (before$\to$after)} &
  \textbf{Loghub 2.0} &
  \textbf{LogBatcher} &
  \multicolumn{1}{l|}{\textbf{UNLEASH}} &
  \textbf{LUNAR} \\ \hline
Multiple Spaces (MS) &
  Replace multiple spaces with a single space. &
  Input:\textcolor{red}{\_\_}\textless$*$\textgreater$\to$Input:\textcolor{cyan}{\_}\textless$*$\textgreater &
  $\sqrt{}$ &
  $\sqrt{}$ &
  $\times$ &
  $\sqrt{}$ \\
Boolean (BL) &
  Replace True/False with a variable. &
  cancel=\textcolor{red}{false}$\to$cancel=\textcolor{cyan}{\textless$*$\textgreater} &
  $\sqrt{}$ &
  $\sqrt{}$ &
  $\times$ &
  $\times$ \\
User-defined String (US) &
  Replace user-defined strings with variables. &
  PCI \textcolor{red}{Root} Bridge$\to$PCI \textcolor{cyan}{\textless$*$\textgreater} Bridge &
  $\sqrt{}$ &
  $\sqrt{}$ &
  $\times$ &
  $\times$ \\
Digit (DG) &
  Replace digit tokens with variables. &
  euid=\textcolor{red}{0}$\to$euid=\textcolor{cyan}{\textless$*$\textgreater} &
  $\sqrt{}$ &
  $\sqrt{}$ &
  $\times$ &
  $\sqrt{}$ \\
Hexadecimal (HEX) &
  Replace hexadecimal numbers with variables. &
  \textcolor{red}{0x1001fdfe} \textless ok\textgreater$\to$\textcolor{cyan}{\textless$*$\textgreater} \textless ok\textgreater &
  $\times$ &
  $\times$ &
  $\times$ &
  $\sqrt{}$ \\
Path-like String (PS) &
  Replace path-like tokens with variables. &
  \textcolor{red}{/lib/tmp} started$\to$\textcolor{cyan}{\textless$*$\textgreater} started &
  $\sqrt{}$ &
  $\sqrt{}$ &
  $\times$ &
  $\times$ \\
Mixed Token (MT) &
  Replace tokens containing variable parts with variables. &
  map-task: \textcolor{red}{'\textless$*$\textgreater'}$\to$map-task: \textcolor{cyan}{\textless$*$\textgreater} &
  $\sqrt{}$ &
  $\sqrt{}$ &
  $\times$ &
  $\sqrt{}$ \\
Delimiter-Separated Variables (DV) &
  Replace delimiter-separated variables as a single variable. &
  Input split: \textcolor{red}{\textless$*$\textgreater+\textless$*$\textgreater}$\to$Input split: \textcolor{cyan}{\textless$*$\textgreater} &
  $\sqrt{}$ &
  $\sqrt{}$ &
  $\sqrt{}$ &
  $\sqrt{}$ \\
Consecutive Variables (CV) &
  Replace consecutive variables as a single variable. &
  value=\textcolor{red}{\textless$*$\textgreater\textless$*$\textgreater}$\to$value=\textcolor{cyan}{\textless$*$\textgreater} &
  $\sqrt{}$ &
  $\sqrt{}$ &
  $\sqrt{}$ &
  $\sqrt{}$ \\
Value Assignments (VA) &
  Replace the ``value'' in a ``key=value'' pair as a variable. &
  tty=\textcolor{red}{ssh}$\to$tty=\textcolor{cyan}{\textless$*$\textgreater} &
  $\times$ &
  $\sqrt{}$ &
  $\times$ &
  $\sqrt{}$ \\ \hline
\end{tabular}%
}
\end{table*}

%% file: sections/6_experiments.tex
\section{Results}
\label{sec:results}

\subsection*{\textbf{RQ1: How do inconsistencies among log data ground-truth versions influence the reliability of parser evaluation results?}}
\label{sec:rq1}

\textbf{The correction rules are implemented differently across versions, and LUNAR's ground-truth version largely differs from the other four versions on both template and message levels.} The two authors examining the template correction descriptions and codes provided an identical categorization of correction rules, as shown in Table~\ref{tab:correction_rules}. Among the 10 correction rules, 1 rule (HEX) is implemented only in LUNAR, 7 rules (BL, US, MS, MT, DG, PS, and VA) are implemented in at least two versions, and the remaining 2 rules (DV and CV) are implemented in all corrected ground-truths. Although the two rules are used in all functions, their implementation details are different. For example, the DS correction in Loghub 2.0 only considers dots (``.'') as delimiters, while other symbols such as dashes (``-'') and colons (``:'') are also treated as delimiters in LUNAR. 

The ratios of events and messages that have different templates are shown in Table~\ref{tab:version_differences}. Given that the correction rules are largely enriched in LUNAR, compared to other versions, it has around 10\% events with different templates, and the difference is always greater than 13\% on the message level. In comparison, the remaining four ground-truth versions exhibit less difference, especially at the message level. 

\textbf{The corrections do not affect GA and FGA. However, parsers' PA and FTA can largely vary across different ground-truth versions, and parsers with better parsing performances are more sensitive to changes.} Although template correction may change grouping labels and evaluation outcomes, the case is not observed in the Loghub 2.0 dataset (i.e., the min-max differences for GA and FGA are always 0). The score shifts on PA and FTA are shown in Table~\ref{tab:supervised_metric_diffs}. We highlighted the differences larger than or equal to 0.04 as the metric score differences are usually within $\pm$ 0.04~\cite{khan2022guidelines}. For all datasets except Proxifier (none of the templates are corrected in this dataset), the impact of template correction can be observed on most of the tools. We also noticed that parsers with the lowest parsing accuracies (i.e., Drain and LibreLog) are less sensitive to the corrections, while the tools with the highest parsing accuracies (i.e., LogBatcher and LUNAR) have many score differences larger than 0.04. Specifically, the min-max score difference on LUNAR's PA is 0.517 on the OpenStack dataset, exhibiting a significant gap. 

\input{tables/version_differences}

\input{tables/supervised_metric_diffs}

\input{tables/optimal_differences}

%\textbf{The FTA score distribution obtained with LUNAR's ground-truth can be statistically different from different versions.} We conducted multiple-pair statistical analysis with Holm-Bonferroni correction using the PA and FTA obtained with different ground-truth versions. While the PA distributions did not show significant differences, the FTA score distribution obtained from LUNAR's ground-truth varied significantly from other versions. As shown in Table~\ref{tab:statistic_analysis}, four out of six log parsers obtained statistically different score distributions when using LUNAR's ground-truth and other versions. Its ground-truth variations can lead to different performance conclusions on the same dataset. 

\textbf{The discrepancy in PA and FTA scores across versions will lead to different optimal tools on the same dataset.} The value shifts caused by different ground-truth versions can lead to varied optimal-performance conclusions on the same dataset. According to Table~\ref{tab:optimal_differences}, 8 datasets have different tools obtaining the highest PA across the five versions, and 7 datasets have disagreements about the tools with the optimal FTA. For instance, LogBatcher has the optimal PA and FTA for HDFS when evaluated with the two Loghub 2.0 ground-truths, while LUNAR achieved the highest scores on the same dataset when the other three versions are used. The optimal tool inconsistencies caused by score value shifts make it difficult to compare parser effectiveness and make selections. 

%Regardless of external constraints such as resource requirements and privacy concerns, a parser with optimal performance scores on the dataset is usually selected to parse the system's future logs. However, t

\input{tables/metric_results_highlighted}

\subsection*{\textbf{RQ2: What is the relationship between PMSS and FGA-FTA?}}
\label{sec:rq2}

%Since PMSS is designed on the template level, we compared the score with the FGA and FTA pairs calculated with the corrected Loghub 2.0 dataset in Table~\ref{tab:metric_outcomes}. 

\textbf{PMSS selects the same optimal parser as FGA in 7 out of the 14 studied datasets, and log parsers achieving the highest PMSS or FGA differ by only 2.1\% on average in terms of FGA.} According to Table~\ref{tab:metric_outcomes}, the tools having the highest FGA also obtained the highest PMSS on half of the datasets. To further understand how the best parser selected by PMSS differs from the best parser selected by FGA, we define the \textit{FGA-PMSS-Gap} metric: $\frac{FGA_{max}-FGA_{PMSS_{max}}}{FGA_{max}}$, quantified by the relative difference between the FGA of the optimal-PMSS parser and the FGA of the optimal-FGA parser. The average \textit{FGA-PMSS-Gap} on all datasets is 2.1\%. As discussed in Sec.~\ref{sec:unsupervised_clustering_analysis}, PMSS shares similar heuristics with log grouping. Therefore, tools that perform best in template-level grouping tend to share similar conclusions under both PMSS and FGA, despite minor numerical differences. 
%\heng{still not clear what the relative difference mean (how you compare two different metrics is not clear (needs a precise definition); the most straighforward way of evalualtion is probably by how the selected best parsers match}, with an average absolute difference of 0.019

\textbf{Given that PMSS is calculated with semantical-structural similarity instead of boolean identicalness, it selects the same optimal parser as FTA on only 3 datasets.} FTA and PMSS both assess template quality in addition to grouping performance. Nevertheless, we observed a larger difference in optimal parser selection between PMSS and FTA. Following the analysis on the FGA-PMSS relationship, we define the \textit{FTA-PMSS-Gap} as $\frac{FTA_{max}-FTA_{PMSS_{max}}}{FTA_{max}}$, and find that the average \textit{FTA-PMSS-Gap} is 9.8\%. This larger gap stems from their different template quality criteria: FTA measures boolean ground-truth identicalness, while PMSS measures semantic–structural similarity. A parser can achieve high PMSS even when its templates are not identical to the ground-truth, as long as they remain semantically and structurally close to their corresponding messages. Conversely, a parser producing fewer but perfectly correct templates may obtain a lower PMSS if its remaining groups show weaker message–template similarity. For example, on ZooKeeper, LogBatcher generated the most ground-truth-identical templates according to FTA, but failed to merge three nearly identical event types into one template:
``Notification: \textless$*$\textgreater (n.leader), \textless$*$\textgreater (n.zxid), \textless$*$\textgreater (n.round), \textless$*$\textgreater (n.state), \textless$*$\textgreater (n.sid), \textless$*$\textgreater (n.peerEPoch), \textless$*$\textgreater (my state)''. Given the templates' high similarity, the EMSS values for these log groups are below 0.2, indicating weak event cohesion. On the other hand, although Preprocessed-Drain generated fewer ground-truth-identical templates, it merged the largest number of highly similar messages, leading to a higher PMSS value. 

\textbf{PMSS shows significant positive correlations with both FGA and FTA.} Considering the small sample size (i.e., 6 log parsers), we did not compare metric rankings per dataset. Instead, we conducted Spearman's $\rho$ analyses with all 84 sample points to understand the correlations between PMSS and FGA-FTA. The correlations of PMSS–FGA and PMSS–FTA are statistically significant ($p<1e^{-8}$), with coefficients of 0.648 and 0.587, respectively. For reference, the correlation between FGA and FTA is 0.670 ($p<1e^{-8}$), indicating that PMSS exhibits a level of alignment with the label-based metrics comparable to the relationship between FGA and FTA themselves. These results suggest that PMSS effectively evaluates log parsers in terms of both event grouping and template quality, reinforcing its potential as a label-free yet integrative proxy for overall parsing performance. 

\textbf{Nevertheless, differences in evaluation strategies lead to some divergent ranking patterns between PMSS, FGA, and FTA.} By design, PMSS highlights parsers achieving balanced and optimal grouping and parsing performance. Following this principle, LUNAR should achieve the highest PMSS on Proxifier, OpenStack, and BGL, while LogBatcher should have the highest PMSS on ZooKeeper and HDFS, as they gained the highest FGA and FTA scores on these datasets. However, neither tool obtained the highest PMSS score. The reasons are as follows. Unlike FTA, which measures identicalness against ground-truth templates, PMSS quantifies similarity, often favoring parsers that produce highly consistent, yet imperfect templates, as previously discussed with the ZooKeeper example. On the other hand, FGA assesses grouping quality purely by predefined event labels, disregarding message–template similarity. Consequently, PMSS may yield rankings that differ from those of both FGA and FTA, reflecting their distinct, yet complementary evaluation perspectives.

\subsection*{\textbf{RQ3: How efficient is the calculation of PMSS in comparison to FGA and FTA?}}
\label{sec:rq3}

%We compare the metric computation time for PMSS to the supervised metrics. The evaluation time for the four supervised metrics is computed with the corrected Loghub 2.0 dataset. The comparison result is illustrated in Figure~\ref{fig:time_consumption}. 

%\begin{figure}[h!]
  %\center
  %\includegraphics[keepaspectratio=1,width=\columnwidth]{figures/time_consumption.pdf}
  %\caption{The total evaluation time of PMSS, FGA, and FTA  for 3 representative tools on different datasets. The remaining 3 tools share a same time consumption trend as Drain. }
  %\label{fig:time_consumption}
%\end{figure}

\input{tables/time_consumption}

\textbf{FGA and FTA generally evaluate faster than PMSS on most datasets.} Due to space constraint, % better visualization, 
we only reported three representative tools' evaluation time consumptions per 1,000 lines in Table~\ref{tab:time_consumption}. The total time consumption of all 6 tools is available in our replication package. FTA compares the identicalness between each pair of parsed and ground-truth templates, and its time consumption per 1,000 lines is generally less than 1ms. On the other hand, FGA compares the size identicalness between each parsed and ground-truth cluster. The process involves filtering each cluster from the parsing result. Thus, FGA requires more evaluation time than FTA, and its time consumption per 1,000 lines fluctuates across datasets. Guided by labels, both metrics compute faster than PMSS in general. PMSS requires additional computations (e.g., regex-based variable filtering) beyond simple equality checks, leading to greater time complexity.

%\textbf{FGA and FTA generally evaluate faster than PMSS on most datasets.} For better visualization, we only printed three representative tool evaluation time consumptions in Figure~\ref{fig:time_consumption}. The time consumption of all 6 tools is available in our replication package. FTA compares the identicalness between each pair of parsed and ground-truth templates, and its evaluation time primarily depends on the number of template pairs. On the other hand, FGA compares the size identicalness between each parsed and ground-truth cluster. The process involves filtering each cluster from the parsing result. Thus, FGA requires more evaluation time than FTA. Guided by labels, both metrics compute faster than PMSS in general. In contrast, PMSS requires additional computations beyond simple equality checks, leading to greater time complexity.
%\heng{be more specific about the algorithm (approach/algorithm matters, implementation less important)}

\textbf{The computation time of PMSS depends on multiple factors but remains near-linear in most cases.} PMSS is calculated per log message, which introduces additional overhead relative to label-based metrics. Its runtime is influenced by factors such as message length, the number of inferred variables, and data preprocessing (e.g., regex matching). As shown in Table~\ref{tab:time_consumption}, four notable peaks in the evaluation time occur for LibreLog on the Mac, BGL and Thunderbird datasets, and for LUNAR on the HPC dataset. These anomalies are due to the existence of unusually long templates and a high volume of inferred variables (i.e., placeholders in templates). 

Despite these anomalies, PMSS's runtime scales near-linearly with the number of log messages, with a general time consumption of around 15ms per 1,000 lines. The label-free evaluation process can be more predictable than FGA. For instance, on the Thunderbird dataset, FGA requires approximately 20ms per 1,000 lines (350 seconds in total), whereas PMSS completes evaluation in approximately 265 seconds. Even with LibreLog considered, PMSS requires less than 300 seconds for computation ($\approx$17ms per 1,000 lines). This relatively consistent scaling ensures that PMSS can assess log parser performance within a reasonable and stable time frame.

%% file: tables/version_differences.tex
% \usepackage{multirow}
% \usepackage{graphicx}
\setlength{\tabcolsep}{2pt}
\begin{table}[]
\centering
\caption{The ratio of events (Temp. Diff.) and messages (Msg. Diff.) having different templates between each version pair. (V1: Original Loghub 2.0, V2: Corrected Loghub 2.0; V3: LogBatcher; V4: UNLEASH; V5: LUNAR)}
\label{tab:version_differences}
\resizebox{\columnwidth}{!}{%
\begin{tabular}{c|llll|lll|ll|l}
\hline
\multirow{2}{*}{\textbf{\begin{tabular}[c]{@{}c@{}}Version\\ Compaison\end{tabular}}} &
  \multicolumn{4}{c|}{\textbf{V1}} &
  \multicolumn{3}{c|}{\textbf{V2}} &
  \multicolumn{2}{c|}{\textbf{V3}} &
  \multicolumn{1}{c}{\textbf{V4}} \\ \cline{2-11} 
 &
  \multicolumn{1}{c|}{\textbf{V2}} &
  \multicolumn{1}{c|}{\textbf{V3}} &
  \multicolumn{1}{c|}{\textbf{V4}} &
  \multicolumn{1}{c|}{\textbf{V5}} &
  \multicolumn{1}{c|}{\textbf{V3}} &
  \multicolumn{1}{c|}{\textbf{V4}} &
  \multicolumn{1}{c|}{\textbf{V5}} &
  \multicolumn{1}{c|}{\textbf{V4}} &
  \multicolumn{1}{c|}{\textbf{V5}} &
  \multicolumn{1}{c}{\textbf{V5}} \\ \hline
\textbf{Temp. Diff. (\%)} &
  \multicolumn{1}{l|}{0.344} &
  \multicolumn{1}{l|}{6.193} &
  \multicolumn{1}{l|}{3.641} &
  12.127 &
  \multicolumn{1}{l|}{5.849} &
  \multicolumn{1}{l|}{3.985} &
  12.443 &
  \multicolumn{1}{l|}{2.724} &
  9.977 &
  9.977 \\
\textbf{Msg. Diff. (\%)} &
  \multicolumn{1}{l|}{0.015} &
  \multicolumn{1}{l|}{1.740} &
  \multicolumn{1}{l|}{1.529} &
  14.528 &
  \multicolumn{1}{l|}{1.725} &
  \multicolumn{1}{l|}{1.544} &
  14.543 &
  \multicolumn{1}{l|}{0.224} &
  13.637 &
  13.568 \\ \hline
\end{tabular}%
}
\end{table}

%% file: tables/supervised_metric_diffs.tex
% \usepackage{multirow}
% \usepackage{graphicx}
\setlength{\tabcolsep}{2pt}
\begin{table}[]
\centering
\caption{The difference between the largest and the smallest values for PA and FTA (i.e., $\Delta$PA and $\Delta$FTA) among all ground-truth versions. Values larger than or equal to 0.04 are highlighted. }
\label{tab:supervised_metric_diffs}
\resizebox{\columnwidth}{!}{%
\begin{tabular}{l|ll|ll|ll|ll|ll|ll}
\hline
\multicolumn{1}{c|}{\multirow{2}{*}{\textbf{Datasets}}} &
  \multicolumn{2}{c|}{\textbf{Drain}} &
  \multicolumn{2}{c|}{\textbf{Pre-Drain}} &
  \multicolumn{2}{c|}{\textbf{LILAC}} &
  \multicolumn{2}{c|}{\textbf{LibreLog}} &
  \multicolumn{2}{c|}{\textbf{LogBatcher}} &
  \multicolumn{2}{c}{\textbf{LUNAR}} \\ \cline{2-13} 
\multicolumn{1}{c|}{} &
  \multicolumn{1}{c|}{\textbf{$\Delta$PA}} &
  \multicolumn{1}{c|}{\textbf{$\Delta$FTA}} &
  \multicolumn{1}{c|}{\textbf{$\Delta$PA}} &
  \multicolumn{1}{c|}{\textbf{$\Delta$FTA}} &
  \multicolumn{1}{c|}{\textbf{$\Delta$PA}} &
  \multicolumn{1}{c|}{\textbf{$\Delta$FTA}} &
  \multicolumn{1}{c|}{\textbf{$\Delta$PA}} &
  \multicolumn{1}{c|}{\textbf{$\Delta$FTA}} &
  \multicolumn{1}{c|}{\textbf{$\Delta$PA}} &
  \multicolumn{1}{c|}{\textbf{$\Delta$FTA}} &
  \multicolumn{1}{c|}{\textbf{$\Delta$PA}} &
  \multicolumn{1}{c}{\textbf{$\Delta$FTA}} \\ \hline
\textbf{Proxifier} &
  \multicolumn{1}{l|}{0.000} &
  0.000 &
  \multicolumn{1}{l|}{0.000} &
  0.000 &
  \multicolumn{1}{l|}{0.000} &
  0.000 &
  \multicolumn{1}{l|}{0.000} &
  0.000 &
  \multicolumn{1}{l|}{0.000} &
  0.000 &
  \multicolumn{1}{l|}{0.000} &
  0.000 \\
\textbf{Linux} &
  \multicolumn{1}{l|}{0.001} &
  0.005 &
  \multicolumn{1}{l|}{0.004} &
  0.035 &
  \multicolumn{1}{l|}{0.015} &
  \textbf{0.061} &
  \multicolumn{1}{l|}{0.002} &
  0.012 &
  \multicolumn{1}{l|}{0.006} &
  \textbf{0.054} &
  \multicolumn{1}{l|}{0.013} &
  \textbf{0.046} \\
\textbf{Apache} &
  \multicolumn{1}{l|}{0.000} &
  0.000 &
  \multicolumn{1}{l|}{0.006} &
  0.034 &
  \multicolumn{1}{l|}{0.025} &
  \textbf{0.069} &
  \multicolumn{1}{l|}{0.024} &
  0.035 &
  \multicolumn{1}{l|}{0.024} &
  0.032 &
  \multicolumn{1}{l|}{0.025} &
  0.035 \\
\textbf{Zookeeper} &
  \multicolumn{1}{l|}{0.001} &
  0.013 &
  \multicolumn{1}{l|}{0.001} &
  0.024 &
  \multicolumn{1}{l|}{0.000} &
  0.023 &
  \multicolumn{1}{l|}{0.000} &
  0.024 &
  \multicolumn{1}{l|}{0.000} &
  0.000 &
  \multicolumn{1}{l|}{0.000} &
  0.024 \\
\textbf{Mac} &
  \multicolumn{1}{l|}{0.024} &
  0.001 &
  \multicolumn{1}{l|}{\textbf{0.040}} &
  0.030 &
  \multicolumn{1}{l|}{\textbf{0.045}} &
  \textbf{0.041} &
  \multicolumn{1}{l|}{0.023} &
  0.017 &
  \multicolumn{1}{l|}{\textbf{0.053}} &
  \textbf{0.083} &
  \multicolumn{1}{l|}{\textbf{0.095}} &
  \textbf{0.103} \\
\textbf{Hadoop} &
  \multicolumn{1}{l|}{0.005} &
  0.010 &
  \multicolumn{1}{l|}{\textbf{0.126}} &
  0.009 &
  \multicolumn{1}{l|}{\textbf{0.131}} &
  \textbf{0.047} &
  \multicolumn{1}{l|}{0.002} &
  0.016 &
  \multicolumn{1}{l|}{\textbf{0.099}} &
  0.020 &
  \multicolumn{1}{l|}{\textbf{0.127}} &
  0.037 \\
\textbf{OpenStack} &
  \multicolumn{1}{l|}{0.009} &
  0.000 &
  \multicolumn{1}{l|}{\textbf{0.467}} &
  \textbf{0.047} &
  \multicolumn{1}{l|}{\textbf{0.042}} &
  \textbf{0.104} &
  \multicolumn{1}{l|}{0.008} &
  0.021 &
  \multicolumn{1}{l|}{\textbf{0.046}} &
  \textbf{0.085} &
  \multicolumn{1}{l|}{\textbf{0.517}} &
  \textbf{0.188} \\
\textbf{HealthApp} &
  \multicolumn{1}{l|}{0.000} &
  0.000 &
  \multicolumn{1}{l|}{0.007} &
  0.032 &
  \multicolumn{1}{l|}{0.015} &
  \textbf{0.057} &
  \multicolumn{1}{l|}{0.004} &
  0.019 &
  \multicolumn{1}{l|}{0.007} &
  0.039 &
  \multicolumn{1}{l|}{0.003} &
  0.026 \\
\textbf{HPC} &
  \multicolumn{1}{l|}{0.000} &
  0.005 &
  \multicolumn{1}{l|}{0.000} &
  0.000 &
  \multicolumn{1}{l|}{0.000} &
  0.000 &
  \multicolumn{1}{l|}{0.000} &
  0.000 &
  \multicolumn{1}{l|}{0.002} &
  0.012 &
  \multicolumn{1}{l|}{0.000} &
  0.012 \\
\textbf{OpenSSH} &
  \multicolumn{1}{l|}{0.000} &
  0.000 &
  \multicolumn{1}{l|}{0.000} &
  0.026 &
  \multicolumn{1}{l|}{0.000} &
  0.030 &
  \multicolumn{1}{l|}{0.000} &
  0.026 &
  \multicolumn{1}{l|}{0.000} &
  0.024 &
  \multicolumn{1}{l|}{0.000} &
  0.026 \\
\textbf{BGL} &
  \multicolumn{1}{l|}{\textbf{0.049}} &
  0.011 &
  \multicolumn{1}{l|}{\textbf{0.063}} &
  0.019 &
  \multicolumn{1}{l|}{\textbf{0.063}} &
  \textbf{0.049} &
  \multicolumn{1}{l|}{\textbf{0.049}} &
  0.016 &
  \multicolumn{1}{l|}{\textbf{0.069}} &
  \textbf{0.047} &
  \multicolumn{1}{l|}{\textbf{0.083}} &
  \textbf{0.054} \\
\textbf{HDFS} &
  \multicolumn{1}{l|}{\textbf{0.052}} &
  \textbf{0.131} &
  \multicolumn{1}{l|}{\textbf{0.052}} &
  \textbf{0.044} &
  \multicolumn{1}{l|}{0.000} &
  \textbf{0.055} &
  \multicolumn{1}{l|}{0.000} &
  0.000 &
  \multicolumn{1}{l|}{\textbf{0.052}} &
  \textbf{0.236} &
  \multicolumn{1}{l|}{\textbf{0.052}} &
  \textbf{0.237} \\
\textbf{Spark} &
  \multicolumn{1}{l|}{\textbf{0.165}} &
  0.040 &
  \multicolumn{1}{l|}{0.002} &
  0.008 &
  \multicolumn{1}{l|}{\textbf{0.282}} &
  \textbf{0.106} &
  \multicolumn{1}{l|}{\textbf{0.164}} &
  0.031 &
  \multicolumn{1}{l|}{\textbf{0.254}} &
  \textbf{0.099} &
  \multicolumn{1}{l|}{\textbf{0.281}} &
  \textbf{0.093} \\
\textbf{Thunderbird} &
  \multicolumn{1}{l|}{0.003} &
  0.001 &
  \multicolumn{1}{l|}{\textbf{0.040}} &
  0.027 &
  \multicolumn{1}{l|}{0.017} &
  0.018 &
  \multicolumn{1}{l|}{0.004} &
  0.021 &
  \multicolumn{1}{l|}{0.014} &
  \textbf{0.043} &
  \multicolumn{1}{l|}{0.014} &
  \textbf{0.055} \\ \hline
\end{tabular}%
}
\end{table}

%% file: tables/optimal_differences.tex
% \usepackage{multirow}
% \usepackage{graphicx}
\begin{table}[]
\centering
\caption{The log parsers with optimal PA and FTA on different ground-truth versions. The dataset-metric pairs with different optimal log parsers are highlighted in gray. }
\label{tab:optimal_differences}
\resizebox{0.8\columnwidth}{!}{%
\begin{tabular}{l|lllll}
\hline
\multicolumn{1}{c|}{\multirow{2}{*}{\textbf{Datasets}}} &
  \multicolumn{5}{c}{\textbf{PA}} \\ \cline{2-6} 
\multicolumn{1}{c|}{} &
  \multicolumn{1}{c|}{\textbf{V1}} &
  \multicolumn{1}{c|}{\textbf{V2}} &
  \multicolumn{1}{c|}{\textbf{V3}} &
  \multicolumn{1}{c|}{\textbf{V4}} &
  \multicolumn{1}{c}{\textbf{V5}} \\ \hline
\textbf{Proxifier} &
  \multicolumn{1}{l|}{LUNAR} &
  \multicolumn{1}{l|}{LUNAR} &
  \multicolumn{1}{l|}{LUNAR} &
  \multicolumn{1}{l|}{LUNAR} &
  LUNAR \\
\textbf{Linux} &
  \multicolumn{1}{l|}{\cellcolor{gray!20}{LILAC}} &
  \multicolumn{1}{l|}{\cellcolor{gray!20}{LILAC}} &
  \multicolumn{1}{l|}{\cellcolor{gray!20}{LILAC}} &
  \multicolumn{1}{l|}{\cellcolor{gray!20}{LILAC}} &
  \cellcolor{gray!20}{LUNAR} \\
\textbf{Apache} &
  \multicolumn{1}{l|}{\cellcolor{gray!20}{LILAC}} &
  \multicolumn{1}{l|}{\cellcolor{gray!20}{LILAC}} &
  \multicolumn{1}{l|}{\cellcolor{gray!20}{LUNAR}} &
  \multicolumn{1}{l|}{\cellcolor{gray!20}{LUNAR}} &
  \cellcolor{gray!20}{LUNAR} \\
\textbf{Zookeeper} &
  \multicolumn{1}{l|}{Drain} &
  \multicolumn{1}{l|}{Drain} &
  \multicolumn{1}{l|}{Drain} &
  \multicolumn{1}{l|}{Drain} &
  Drain \\
\textbf{Mac} &
  \multicolumn{1}{l|}{\cellcolor{gray!20}{LogBatcher}} &
  \multicolumn{1}{l|}{\cellcolor{gray!20}{LogBatcher}} &
  \multicolumn{1}{l|}{\cellcolor{gray!20}{LogBatcher}} &
  \multicolumn{1}{l|}{\cellcolor{gray!20}{LogBatcher}} &
  \cellcolor{gray!20}{LUNAR} \\
\textbf{Hadoop} &
  \multicolumn{1}{l|}{\cellcolor{gray!20}{LILAC}} &
  \multicolumn{1}{l|}{\cellcolor{gray!20}{LILAC}} &
  \multicolumn{1}{l|}{\cellcolor{gray!20}{LILAC}} &
  \multicolumn{1}{l|}{\cellcolor{gray!20}{LILAC}} &
  \cellcolor{gray!20}{LUNAR} \\
\textbf{OpenStack} &
  \multicolumn{1}{l|}{\cellcolor{gray!20}{LUNAR}} &
  \multicolumn{1}{l|}{\cellcolor{gray!20}{LUNAR}} &
  \multicolumn{1}{l|}{\cellcolor{gray!20}{Pre-Drain}} &
  \multicolumn{1}{l|}{\cellcolor{gray!20}{Pre-Drain}} &
  \cellcolor{gray!20}{LUNAR} \\
\textbf{HealthApp} &
  \multicolumn{1}{l|}{LUNAR} &
  \multicolumn{1}{l|}{LUNAR} &
  \multicolumn{1}{l|}{LUNAR} &
  \multicolumn{1}{l|}{LUNAR} &
  LUNAR \\
\textbf{HPC} &
  \multicolumn{1}{l|}{LUNAR} &
  \multicolumn{1}{l|}{LUNAR} &
  \multicolumn{1}{l|}{LUNAR} &
  \multicolumn{1}{l|}{LUNAR} &
  LUNAR \\
\textbf{OpenSSH} &
  \multicolumn{1}{l|}{LUNAR} &
  \multicolumn{1}{l|}{LUNAR} &
  \multicolumn{1}{l|}{LUNAR} &
  \multicolumn{1}{l|}{LUNAR} &
  LUNAR \\
\textbf{BGL} &
  \multicolumn{1}{l|}{\cellcolor{gray!20}{LogBatcher}} &
  \multicolumn{1}{l|}{\cellcolor{gray!20}{LogBatcher}} &
  \multicolumn{1}{l|}{\cellcolor{gray!20}{LogBatcher}} &
  \multicolumn{1}{l|}{\cellcolor{gray!20}{LogBatcher}} &
  \cellcolor{gray!20}{LUNAR} \\
\textbf{HDFS} &
  \multicolumn{1}{l|}{\cellcolor{gray!20}{LogBatcher}} &
  \multicolumn{1}{l|}{\cellcolor{gray!20}{LogBatcher}} &
  \multicolumn{1}{l|}{\cellcolor{gray!20}{LUNAR}} &
  \multicolumn{1}{l|}{\cellcolor{gray!20}{LUNAR}} &
  \cellcolor{gray!20}{LUNAR} \\
\textbf{Spark} &
  \multicolumn{1}{l|}{\cellcolor{gray!20}{LILAC}} &
  \multicolumn{1}{l|}{\cellcolor{gray!20}{LILAC}} &
  \multicolumn{1}{l|}{\cellcolor{gray!20}{LILAC}} &
  \multicolumn{1}{l|}{\cellcolor{gray!20}{LILAC}} &
  \cellcolor{gray!20}{LUNAR} \\
\textbf{Thunderbird} &
  \multicolumn{1}{l|}{LUNAR} &
  \multicolumn{1}{l|}{LUNAR} &
  \multicolumn{1}{l|}{LUNAR} &
  \multicolumn{1}{l|}{LUNAR} &
  LUNAR \\ \hline
\multicolumn{1}{c|}{\multirow{2}{*}{\textbf{Datasets}}} &
  \multicolumn{5}{c}{\textbf{FTA}} \\ \cline{2-6} 
\multicolumn{1}{c|}{} &
  \multicolumn{1}{c|}{\textbf{V1}} &
  \multicolumn{1}{c|}{\textbf{V2}} &
  \multicolumn{1}{c|}{\textbf{V3}} &
  \multicolumn{1}{c|}{\textbf{V4}} &
  \multicolumn{1}{c}{\textbf{V5}} \\ \hline
\textbf{Proxifier} &
  \multicolumn{1}{l|}{LUNAR} &
  \multicolumn{1}{l|}{LUNAR} &
  \multicolumn{1}{l|}{LUNAR} &
  \multicolumn{1}{l|}{LUNAR} &
  LUNAR \\
\textbf{Linux} &
  \multicolumn{1}{l|}{\cellcolor{gray!20}{LogBatcher}} &
  \multicolumn{1}{l|}{\cellcolor{gray!20}{LogBatcher}} &
  \multicolumn{1}{l|}{\cellcolor{gray!20}{LUNAR}} &
  \multicolumn{1}{l|}{\cellcolor{gray!20}{LogBatcher}} &
  \cellcolor{gray!20}{LUNAR} \\
\textbf{Apache} &
  \multicolumn{1}{l|}{LUNAR} &
  \multicolumn{1}{l|}{LUNAR} &
  \multicolumn{1}{l|}{LUNAR} &
  \multicolumn{1}{l|}{LUNAR} &
  LUNAR \\
\textbf{Zookeeper} &
  \multicolumn{1}{l|}{LogBatcher} &
  \multicolumn{1}{l|}{LogBatcher} &
  \multicolumn{1}{l|}{LogBatcher} &
  \multicolumn{1}{l|}{LogBatcher} &
  LogBatcher \\
\textbf{Mac} &
  \multicolumn{1}{l|}{\cellcolor{gray!20}{LogBatcher}} &
  \multicolumn{1}{l|}{\cellcolor{gray!20}{LogBatcher}} &
  \multicolumn{1}{l|}{\cellcolor{gray!20}{LUNAR}} &
  \multicolumn{1}{l|}{\cellcolor{gray!20}{LUNAR}} &
  \cellcolor{gray!20}{LUNAR} \\
\textbf{Hadoop} &
  \multicolumn{1}{l|}{\cellcolor{gray!20}{LILAC}} &
  \multicolumn{1}{l|}{\cellcolor{gray!20}{LILAC}} &
  \multicolumn{1}{l|}{\cellcolor{gray!20}{LILAC}} &
  \multicolumn{1}{l|}{\cellcolor{gray!20}{LILAC}} &
  \cellcolor{gray!20}{LUNAR} \\
\textbf{OpenStack} &
  \multicolumn{1}{l|}{\cellcolor{gray!20}{LUNAR}} &
  \multicolumn{1}{l|}{\cellcolor{gray!20}{LUNAR}} &
  \multicolumn{1}{l|}{\cellcolor{gray!20}{LogBatcher}} &
  \multicolumn{1}{l|}{\cellcolor{gray!20}{LILAC}} &
  \cellcolor{gray!20}{LUNAR} \\
\textbf{HealthApp} &
  \multicolumn{1}{l|}{LUNAR} &
  \multicolumn{1}{l|}{LUNAR} &
  \multicolumn{1}{l|}{LUNAR} &
  \multicolumn{1}{l|}{LUNAR} &
  LUNAR \\
\textbf{HPC} &
  \multicolumn{1}{l|}{LUNAR} &
  \multicolumn{1}{l|}{LUNAR} &
  \multicolumn{1}{l|}{LUNAR} &
  \multicolumn{1}{l|}{LUNAR} &
  LUNAR \\
\textbf{OpenSSH} &
  \multicolumn{1}{l|}{LUNAR} &
  \multicolumn{1}{l|}{LUNAR} &
  \multicolumn{1}{l|}{LUNAR} &
  \multicolumn{1}{l|}{LUNAR} &
  LUNAR \\
\textbf{BGL} &
  \multicolumn{1}{l|}{LUNAR} &
  \multicolumn{1}{l|}{LUNAR} &
  \multicolumn{1}{l|}{LUNAR} &
  \multicolumn{1}{l|}{LUNAR} &
  LUNAR \\
\textbf{HDFS} &
  \multicolumn{1}{l|}{\cellcolor{gray!20}{LogBatcher}} &
  \multicolumn{1}{l|}{\cellcolor{gray!20}{LogBatcher}} &
  \multicolumn{1}{l|}{\cellcolor{gray!20}{LUNAR}} &
  \multicolumn{1}{l|}{\cellcolor{gray!20}{LUNAR}} &
  \cellcolor{gray!20}{LUNAR} \\
\textbf{Spark} &
  \multicolumn{1}{l|}{\cellcolor{gray!20}{LILAC}} &
  \multicolumn{1}{l|}{\cellcolor{gray!20}{LILAC}} &
  \multicolumn{1}{l|}{\cellcolor{gray!20}{LILAC}} &
  \multicolumn{1}{l|}{\cellcolor{gray!20}{LILAC}} &
  \cellcolor{gray!20}{LUNAR} \\
\textbf{Thunderbird} &
  \multicolumn{1}{l|}{\cellcolor{gray!20}{LogBatcher}} &
  \multicolumn{1}{l|}{\cellcolor{gray!20}{LogBatcher}} &
  \multicolumn{1}{l|}{\cellcolor{gray!20}{LUNAR}} &
  \multicolumn{1}{l|}{\cellcolor{gray!20}{LogBatcher}} &
  \cellcolor{gray!20}{LUNAR} \\ \hline
\end{tabular}%
}
\end{table}

%% file: tables/metric_results_highlighted.tex
% \usepackage{multirow}
% \usepackage{graphicx}
\begin{table*}[]
\centering
\caption{The FGA,  FTA, and PMSS of each tool on 14 datasets. The best values on each dataset are highlighted in bold, and the worst values are highlighted in italics.}%\heng{I think the most important conclusion of this RQ is: PMSS select the same optimal parser as FGA for 7 (?) out of the 14 datasets, or other ways of direclty showing how PMSS is similar to FGA or FTA in terms of parsing selection, like: the best parser selected by PMSS is the best or the second-best parser selected by FGA or FTA in X out of the 14 datasets.}
\label{tab:metric_outcomes}
\resizebox{0.8\textwidth}{!}{%
\begin{tabular}{l|lll|lll|lll|lll|lll|lll}
\hline
\multirow{2}{*}{\textbf{Datasets}} &
  \multicolumn{3}{c|}{\textbf{Drain}} &
  \multicolumn{3}{c|}{\textbf{Pre-Drain}} &
  \multicolumn{3}{c|}{\textbf{LILAC}} &
  \multicolumn{3}{c|}{\textbf{LibreLog}} &
  \multicolumn{3}{c|}{\textbf{LogBatcher}} &
  \multicolumn{3}{c}{\textbf{LUNAR}} \\ \cline{2-19} 
 &
  \multicolumn{1}{l|}{\textbf{FGA}} &
  \multicolumn{1}{l|}{\textbf{FTA}} &
  \textbf{PMSS} &
  \multicolumn{1}{l|}{\textbf{FGA}} &
  \multicolumn{1}{l|}{\textbf{FTA}} &
  \textbf{PMSS} &
  \multicolumn{1}{l|}{\textbf{FGA}} &
  \multicolumn{1}{l|}{\textbf{FTA}} &
  \textbf{PMSS} &
  \multicolumn{1}{l|}{\textbf{FGA}} &
  \multicolumn{1}{l|}{\textbf{FTA}} &
  \textbf{PMSS} &
  \multicolumn{1}{l|}{\textbf{FGA}} &
  \multicolumn{1}{l|}{\textbf{FTA}} &
  \textbf{PMSS} &
  \multicolumn{1}{l|}{\textbf{FGA}} &
  \multicolumn{1}{l|}{\textbf{FTA}} &
  \textbf{PMSS} \\ \hline
\textbf{Proxifier} &
  \multicolumn{1}{l|}{0.206} &
  \multicolumn{1}{l|}{0.176} &
  0.126 &
  \multicolumn{1}{l|}{\textbf{0.870}} &
  \multicolumn{1}{l|}{0.870} &
  \textbf{0.513} &
  \multicolumn{1}{l|}{\textit{0.000}} &
  \multicolumn{1}{l|}{0.065} &
  0.227 &
  \multicolumn{1}{l|}{0.208} &
  \multicolumn{1}{l|}{\textit{0.000}} &
  \textit{0.033} &
  \multicolumn{1}{l|}{0.762} &
  \multicolumn{1}{l|}{0.762} &
  0.464 &
  \multicolumn{1}{l|}{\textbf{0.870}} &
  \multicolumn{1}{l|}{\textbf{0.957}} &
  0.501 \\
\textbf{Linux} &
  \multicolumn{1}{l|}{0.778} &
  \multicolumn{1}{l|}{\textit{0.257}} &
  0.719 &
  \multicolumn{1}{l|}{0.857} &
  \multicolumn{1}{l|}{0.524} &
  0.684 &
  \multicolumn{1}{l|}{0.819} &
  \multicolumn{1}{l|}{0.508} &
  0.666 &
  \multicolumn{1}{l|}{\textit{0.724}} &
  \multicolumn{1}{l|}{0.294} &
  \textit{0.575} &
  \multicolumn{1}{l|}{\textbf{0.894}} &
  \multicolumn{1}{l|}{\textbf{0.696}} &
  \textbf{0.732} &
  \multicolumn{1}{l|}{0.870} &
  \multicolumn{1}{l|}{0.668} &
  0.710 \\
\textbf{Apache} &
  \multicolumn{1}{l|}{\textbf{1.000}} &
  \multicolumn{1}{l|}{\textit{0.517}} &
  0.790 &
  \multicolumn{1}{l|}{0.949} &
  \multicolumn{1}{l|}{0.678} &
  0.769 &
  \multicolumn{1}{l|}{\textbf{1.000}} &
  \multicolumn{1}{l|}{0.690} &
  0.740 &
  \multicolumn{1}{l|}{\textbf{1.000}} &
  \multicolumn{1}{l|}{0.552} &
  \textit{0.657} &
  \multicolumn{1}{l|}{\textit{0.918}} &
  \multicolumn{1}{l|}{0.721} &
  \textbf{0.814} &
  \multicolumn{1}{l|}{\textbf{1.000}} &
  \multicolumn{1}{l|}{\textbf{0.862}} &
  \textbf{0.814} \\
\textbf{Zookeeper} &
  \multicolumn{1}{l|}{0.904} &
  \multicolumn{1}{l|}{0.614} &
  0.817 &
  \multicolumn{1}{l|}{0.904} &
  \multicolumn{1}{l|}{0.747} &
  \textbf{0.818} &
  \multicolumn{1}{l|}{\textit{0.867}} &
  \multicolumn{1}{l|}{0.601} &
  0.686 &
  \multicolumn{1}{l|}{0.897} &
  \multicolumn{1}{l|}{\textit{0.521}} &
  \textit{0.681} &
  \multicolumn{1}{l|}{\textbf{0.908}} &
  \multicolumn{1}{l|}{\textbf{0.828}} &
  0.783 &
  \multicolumn{1}{l|}{0.885} &
  \multicolumn{1}{l|}{0.752} &
  0.802 \\
\textbf{Mac} &
  \multicolumn{1}{l|}{\textit{0.229}} &
  \multicolumn{1}{l|}{\textit{0.069}} &
  \textit{0.148} &
  \multicolumn{1}{l|}{0.818} &
  \multicolumn{1}{l|}{0.394} &
  0.630 &
  \multicolumn{1}{l|}{0.770} &
  \multicolumn{1}{l|}{0.434} &
  0.619 &
  \multicolumn{1}{l|}{0.792} &
  \multicolumn{1}{l|}{0.240} &
  0.352 &
  \multicolumn{1}{l|}{0.869} &
  \multicolumn{1}{l|}{\textbf{0.526}} &
  0.672 &
  \multicolumn{1}{l|}{\textbf{0.873}} &
  \multicolumn{1}{l|}{0.444} &
  \textbf{0.673} \\
\textbf{Hadoop} &
  \multicolumn{1}{l|}{\textit{0.785}} &
  \multicolumn{1}{l|}{0.384} &
  0.530 &
  \multicolumn{1}{l|}{\textbf{0.947}} &
  \multicolumn{1}{l|}{0.589} &
  0.736 &
  \multicolumn{1}{l|}{0.921} &
  \multicolumn{1}{l|}{\textbf{0.682}} &
  0.715 &
  \multicolumn{1}{l|}{0.913} &
  \multicolumn{1}{l|}{\textit{0.371}} &
  \textit{0.481} &
  \multicolumn{1}{l|}{0.845} &
  \multicolumn{1}{l|}{0.638} &
  0.711 &
  \multicolumn{1}{l|}{0.915} &
  \multicolumn{1}{l|}{0.635} &
  \textbf{0.768} \\
\textbf{OpenStack} &
  \multicolumn{1}{l|}{\textit{0.007}} &
  \multicolumn{1}{l|}{\textit{0.002}} &
  \textit{0.125} &
  \multicolumn{1}{l|}{0.279} &
  \multicolumn{1}{l|}{0.163} &
  0.358 &
  \multicolumn{1}{l|}{\textbf{1.000}} &
  \multicolumn{1}{l|}{0.833} &
  0.614 &
  \multicolumn{1}{l|}{0.653} &
  \multicolumn{1}{l|}{0.168} &
  0.308 &
  \multicolumn{1}{l|}{0.957} &
  \multicolumn{1}{l|}{0.723} &
  \textbf{0.680} &
  \multicolumn{1}{l|}{\textbf{1.000}} &
  \multicolumn{1}{l|}{\textbf{0.854}} &
  0.674 \\
\textbf{HealthApp} &
  \multicolumn{1}{l|}{\textit{0.010}} &
  \multicolumn{1}{l|}{\textit{0.004}} &
  \textit{0.003} &
  \multicolumn{1}{l|}{0.946} &
  \multicolumn{1}{l|}{0.629} &
  0.691 &
  \multicolumn{1}{l|}{\textbf{0.971}} &
  \multicolumn{1}{l|}{0.768} &
  0.680 &
  \multicolumn{1}{l|}{0.958} &
  \multicolumn{1}{l|}{0.581} &
  0.580 &
  \multicolumn{1}{l|}{0.945} &
  \multicolumn{1}{l|}{0.804} &
  0.697 &
  \multicolumn{1}{l|}{\textbf{0.971}} &
  \multicolumn{1}{l|}{\textbf{0.837}} &
  \textbf{0.711} \\
\textbf{HPC} &
  \multicolumn{1}{l|}{\textit{0.309}} &
  \multicolumn{1}{l|}{\textit{0.152}} &
  \textit{0.309} &
  \multicolumn{1}{l|}{0.584} &
  \multicolumn{1}{l|}{0.565} &
  0.411 &
  \multicolumn{1}{l|}{\textbf{0.857}} &
  \multicolumn{1}{l|}{0.586} &
  0.678 &
  \multicolumn{1}{l|}{0.670} &
  \multicolumn{1}{l|}{0.462} &
  0.421 &
  \multicolumn{1}{l|}{0.830} &
  \multicolumn{1}{l|}{0.767} &
  \textbf{0.746} &
  \multicolumn{1}{l|}{0.843} &
  \multicolumn{1}{l|}{\textbf{0.819}} &
  0.689 \\
\textbf{OpenSSH} &
  \multicolumn{1}{l|}{0.872} &
  \multicolumn{1}{l|}{0.487} &
  0.613 &
  \multicolumn{1}{l|}{0.883} &
  \multicolumn{1}{l|}{0.883} &
  0.656 &
  \multicolumn{1}{l|}{\textit{0.727}} &
  \multicolumn{1}{l|}{0.394} &
  0.636 &
  \multicolumn{1}{l|}{0.907} &
  \multicolumn{1}{l|}{\textit{0.347}} &
  \textit{0.423} &
  \multicolumn{1}{l|}{0.840} &
  \multicolumn{1}{l|}{0.765} &
  0.657 &
  \multicolumn{1}{l|}{\textbf{0.923}} &
  \multicolumn{1}{l|}{\textbf{0.897}} &
  \textbf{0.679} \\
\textbf{BGL} &
  \multicolumn{1}{l|}{\textit{0.624}} &
  \multicolumn{1}{l|}{\textit{0.193}} &
  0.550 &
  \multicolumn{1}{l|}{0.759} &
  \multicolumn{1}{l|}{0.601} &
  \textbf{0.674} &
  \multicolumn{1}{l|}{0.820} &
  \multicolumn{1}{l|}{0.631} &
  0.605 &
  \multicolumn{1}{l|}{0.764} &
  \multicolumn{1}{l|}{0.488} &
  \textit{0.482} &
  \multicolumn{1}{l|}{0.827} &
  \multicolumn{1}{l|}{0.716} &
  0.634 &
  \multicolumn{1}{l|}{\textbf{0.872}} &
  \multicolumn{1}{l|}{\textbf{0.738}} &
  0.651 \\
\textbf{HDFS} &
  \multicolumn{1}{l|}{0.935} &
  \multicolumn{1}{l|}{0.609} &
  0.660 &
  \multicolumn{1}{l|}{0.901} &
  \multicolumn{1}{l|}{0.440} &
  0.604 &
  \multicolumn{1}{l|}{\textit{0.658}} &
  \multicolumn{1}{l|}{0.548} &
  0.608 &
  \multicolumn{1}{l|}{0.809} &
  \multicolumn{1}{l|}{\textit{0.106}} &
  \textit{0.382} &
  \multicolumn{1}{l|}{\textbf{0.968}} &
  \multicolumn{1}{l|}{\textbf{0.946}} &
  0.655 &
  \multicolumn{1}{l|}{\textbf{0.968}} &
  \multicolumn{1}{l|}{0.731} &
  \textbf{0.664} \\
\textbf{Spark} &
  \multicolumn{1}{l|}{0.861} &
  \multicolumn{1}{l|}{0.412} &
  0.689 &
  \multicolumn{1}{l|}{0.866} &
  \multicolumn{1}{l|}{0.540} &
  \textbf{0.736} &
  \multicolumn{1}{l|}{0.885} &
  \multicolumn{1}{l|}{0.674} &
  0.685 &
  \multicolumn{1}{l|}{0.788} &
  \multicolumn{1}{l|}{\textit{0.337}} &
  \textit{0.472} &
  \multicolumn{1}{l|}{\textit{0.777}} &
  \multicolumn{1}{l|}{\textbf{0.674}} &
  0.654 &
  \multicolumn{1}{l|}{\textbf{0.888}} &
  \multicolumn{1}{l|}{0.572} &
  0.732 \\
\textbf{Thunderbird} &
  \multicolumn{1}{l|}{\textit{0.237}} &
  \multicolumn{1}{l|}{\textit{0.071}} &
  \textit{0.292} &
  \multicolumn{1}{l|}{0.844} &
  \multicolumn{1}{l|}{0.495} &
  \textbf{0.756} &
  \multicolumn{1}{l|}{0.322} &
  \multicolumn{1}{l|}{0.193} &
  0.294 &
  \multicolumn{1}{l|}{0.843} &
  \multicolumn{1}{l|}{0.282} &
  0.474 &
  \multicolumn{1}{l|}{0.818} &
  \multicolumn{1}{l|}{\textbf{0.566}} &
  0.701 &
  \multicolumn{1}{l|}{\textbf{0.871}} &
  \multicolumn{1}{l|}{0.538} &
  0.714 \\ \hline
\end{tabular}%
}
\end{table*}

%% file: tables/time_consumption.tex
% \usepackage{multirow}
% \usepackage{graphicx}
\begin{table}[]
\centering
\caption{The evaluation time (ms) of PMSS, FGA, and FTA per 1,000 lines for 3 representative tools on different datasets. The remaining 3 tools share the same trend as Drain.}
\label{tab:time_consumption}
\resizebox{\columnwidth}{!}{%
\begin{tabular}{l|lll|lll|lll}
\hline
\multirow{2}{*}{\textbf{Datasets}} &
  \multicolumn{3}{c|}{\textbf{FGA (ms/1k lines)}} &
  \multicolumn{3}{c|}{\textbf{FTA (ms/1k lines)}} &
  \multicolumn{3}{c}{\textbf{PMSS (ms/1k lines)}} \\ \cline{2-10} 
 &
  \multicolumn{1}{l|}{\textbf{Drain}} &
  \multicolumn{1}{l|}{\textbf{LibreLog}} &
  \textbf{LUNAR} &
  \multicolumn{1}{l|}{\textbf{Drain}} &
  \multicolumn{1}{l|}{\textbf{LibreLog}} &
  \textbf{LUNAR} &
  \multicolumn{1}{l|}{\textbf{Drain}} &
  \multicolumn{1}{l|}{\textbf{LibreLog}} &
  \textbf{LUNAR} \\ \hline
\textbf{Proxifier} &
  \multicolumn{1}{l|}{2.81} &
  \multicolumn{1}{l|}{0.94} &
  0.94 &
  \multicolumn{1}{l|}{0.47} &
  \multicolumn{1}{l|}{{0.47}} &
  {0.47} &
  \multicolumn{1}{l|}{16.28} &
  \multicolumn{1}{l|}{20.45} &
  18.39 \\
\textbf{Linux} &
  \multicolumn{1}{l|}{6.27} &
  \multicolumn{1}{l|}{{6.69}} &
  7.11 &
  \multicolumn{1}{l|}{{0.42}} &
  \multicolumn{1}{l|}{0.42} &
  {0.42} &
  \multicolumn{1}{l|}{15.17} &
  \multicolumn{1}{l|}{14.72} &
  16.30 \\
\textbf{Apache} &
  \multicolumn{1}{l|}{0.77} &
  \multicolumn{1}{l|}{{0.77}} &
  0.77 &
  \multicolumn{1}{l|}{0.19} &
  \multicolumn{1}{l|}{0.19} &
  {0.19} &
  \multicolumn{1}{l|}{14.53} &
  \multicolumn{1}{l|}{15.12} &
  14.64 \\
\textbf{Zookeeper} &
  \multicolumn{1}{l|}{1.75} &
  \multicolumn{1}{l|}{1.75} &
  1.75 &
  \multicolumn{1}{l|}{0.27} &
  \multicolumn{1}{l|}{{0.13}} &
  {0.27} &
  \multicolumn{1}{l|}{14.97} &
  \multicolumn{1}{l|}{14.80} &
  14.51 \\
\textbf{Mac} &
  \multicolumn{1}{l|}{9.27} &
  \multicolumn{1}{l|}{{10.17}} &
  {10.67} &
  \multicolumn{1}{l|}{0.80} &
  \multicolumn{1}{l|}{0.40} &
  0.30 &
  \multicolumn{1}{l|}{21.63} &
  \multicolumn{1}{l|}{306.68} &
  17.48 \\
\textbf{Hadoop} &
  \multicolumn{1}{l|}{3.89} &
  \multicolumn{1}{l|}{4.17} &
  4.00 &
  \multicolumn{1}{l|}{0.22} &
  \multicolumn{1}{l|}{{0.22}} &
  {0.22} &
  \multicolumn{1}{l|}{15.12} &
  \multicolumn{1}{l|}{15.35} &
  16.05 \\
\textbf{OpenStack} &
  \multicolumn{1}{l|}{0.72} &
  \multicolumn{1}{l|}{{0.96}} &
  {1.06} &
  \multicolumn{1}{l|}{0.67} &
  \multicolumn{1}{l|}{0.24} &
  0.24 &
  \multicolumn{1}{l|}{22.97} &
  \multicolumn{1}{l|}{15.64} &
  15.74 \\
\textbf{HealthApp} &
  \multicolumn{1}{l|}{{2.68}} &
  \multicolumn{1}{l|}{{2.87}} &
  {2.87} &
  \multicolumn{1}{l|}{1.88} &
  \multicolumn{1}{l|}{0.19} &
  0.19 &
  \multicolumn{1}{l|}{21.90} &
  \multicolumn{1}{l|}{14.74} &
  14.34 \\
\textbf{HPC} &
  \multicolumn{1}{l|}{1.35} &
  \multicolumn{1}{l|}{{1.33}} &
  {1.40} &
  \multicolumn{1}{l|}{0.23} &
  \multicolumn{1}{l|}{0.19} &
  0.21 &
  \multicolumn{1}{l|}{13.80} &
  \multicolumn{1}{l|}{13.91} &
  75.63 \\
\textbf{OpenSSH} &
  \multicolumn{1}{l|}{0.85} &
  \multicolumn{1}{l|}{0.88} &
  0.83 &
  \multicolumn{1}{l|}{0.20} &
  \multicolumn{1}{l|}{{0.19}} &
  0.20 &
  \multicolumn{1}{l|}{15.22} &
  \multicolumn{1}{l|}{15.94} &
  16.15 \\
\textbf{BGL} &
  \multicolumn{1}{l|}{5.30} &
  \multicolumn{1}{l|}{{5.14}} &
  5.24 &
  \multicolumn{1}{l|}{0.29} &
  \multicolumn{1}{l|}{0.20} &
  0.20 &
  \multicolumn{1}{l|}{14.52} &
  \multicolumn{1}{l|}{36.10} &
  14.69 \\
\textbf{HDFS} &
  \multicolumn{1}{l|}{1.27} &
  \multicolumn{1}{l|}{1.11} &
  1.17 &
  \multicolumn{1}{l|}{0.39} &
  \multicolumn{1}{l|}{{0.23}} &
  0.31 &
  \multicolumn{1}{l|}{15.75} &
  \multicolumn{1}{l|}{15.94} &
  15.47 \\
\textbf{Spark} &
  \multicolumn{1}{l|}{4.29} &
  \multicolumn{1}{l|}{3.86} &
  4.15 &
  \multicolumn{1}{l|}{0.38} &
  \multicolumn{1}{l|}{{0.28}} &
  0.29 &
  \multicolumn{1}{l|}{16.19} &
  \multicolumn{1}{l|}{15.85} &
  16.05 \\
\textbf{Thunderbird} &
  \multicolumn{1}{l|}{19.79} &
  \multicolumn{1}{l|}{{19.78}} &
  {20.23} &
  \multicolumn{1}{l|}{0.41} &
  \multicolumn{1}{l|}{0.24} &
  0.26 &
  \multicolumn{1}{l|}{15.99} &
  \multicolumn{1}{l|}{17.29} &
  15.91 \\ \hline
\end{tabular}%
}
\end{table}

\vspace{-2mm}

%% file: sections/7_discussions.tex
\section{Discussions}
\label{sec:discussions}

\subsection{Consideration on Evaluation Levels}
\label{sec:evaluation_levels}

Khan et al.~\cite{khan2022guidelines} distinguished between two levels of parser evaluation: \textbf{message-level metrics} (e.g., GA, PA) measure the parser’s overall performance across all log messages, while \textbf{template-level metrics} (e.g., FGA, FTA) assess the parsing quality of the extracted event clusters. Following this definition, PMSS falls into the category of template-level metrics.

We did not define metrics on the message level because, in comparison to template-level metrics, message-level template-quality metrics are less practical. According to the large-scale survey of He et al.~\cite{he2021survey}, the main need for accurate log grouping and high template quality arises in downstream task automation and sometimes manual inspection. Practitioners rarely examine templates on a per-message basis, since the vast amount of log data makes such inspection infeasible. Instead, they focus on reviewing the representative event templates. Moreover, Khan et al. noted a critical limitation of message-level metrics: they may overlook cases where rare but critical events (e.g., anomalies) are poorly summarized, even though these events are usually important in downstream analysis~\cite{khan2022guidelines}. 

\subsection{Label-free vs. Label-based Metrics}
\label{sec:metric_comparison}

Similar to the four widely used label-based effectiveness metrics (i.e., GA, PA, FGA and FTA), PMSS also evaluates the log parsers' performance on log grouping and template abstraction. However, unlike these metrics, PMSS \textbf{does not rely on the notion of parsing correctness but evaluates event cohesion and separation based on semantic-structural similarities}, as the correctness of parsed templates is often unverifiable due to the absence of logging statements. 

As evaluated in RQ2, PMSS is significantly and positively correlated with both FGA and FTA. The result corresponds to our prior analysis in Sec.~\ref{sec:unsupervised_clustering_analysis} that a single cluster analysis metric (i.e., PMSS) can capture both grouping and template quality of a log parser. Further, we discovered that the tools with the highest PMSS also have a small performance difference from the optimal in terms of FGA. The two metrics share the same evaluation heuristic: log messages should be highly cohesive within events, while the events are well separated. 

However, we found that the highest PMSS does not guarantee a (near-)optimal FTA. This is because FTA uses predefined ground-truth sets to provide boolean justifications on the ``correctness'' of templates. In contrast, PMSS's template evaluation is based on template-message semantic-structural similarity, quantified with the Levenshtein distance between processed templates and messages. An optimal FTA suggests that most templates are \textbf{identical} between the parsing results and ground-truths, whereas an optimal PMSS indicates that most templates share \textbf{highly similar} semantic-structural features with their corresponding messages. %Given the difference in evaluation strategy, their results should be interpreted differently. 

Given their different evaluation strategy, \textbf{PMSS cannot replace the explanation of label-based metrics in performance comprehension}. Further, PMSS's evaluation outcome depends on the data processing approaches used. Specifically, the challenges lie in tokenization (i.e., not all tokens with different semantics are separated by white space) and variable filtering in templates (e.g., the non-numeric variables and variables with multiple tokens). Although PMSS is not identical to label-based metrics, it serves as a powerful alternative when ground-truth labels are inconsistent or unavailable. On the other hand, to mitigate the previously spotted issues in label-based log parser evaluation, we encourage researchers to explicitly state the corrections made to the ground-truth and compare the outcomes obtained from different correction versions to avoid drawing misleading conclusions.

\subsection{Selecting Parsers for Downstream Tasks}
\label{sec:parser_selection}

Log parsing serves as an intermediate step in log analysis: log parsers convert messages into event templates to facilitate downstream log mining tasks, such as anomaly detection, failure diagnosis, and failure prediction~\cite{he2021survey}. Hence, the selection of log parsers should be task-oriented. 

According to He et al.~\cite{he2021survey}, failure diagnosis and failure prediction tools often integrate event sequences (e.g., a trail of event IDs) into their model. Their requirement for event separation is naturally aligned with the evaluation target of PMSS and FGA. On the other hand, more event feature abstraction strategies are used in anomaly detection tools, such as word embeddings or graphical features, which seemingly stress the need of ``correct'' event templates. Nonetheless, a recent study by Khan et al. analyzed the impact of log parsing on anomaly detection~\cite{khan2024impact} and discovered that GA, PA, and FTA do not have a significant correlation with anomaly detection accuracy, regardless of the detection algorithms used. Instead, anomaly detection accuracy shows correlation with distinguishability: i.e., whether the templates of normal and abnormal logs are different. The notion of distinguishability is similar to that of FGA and PMSS, as both metrics stress the separation of logs denoting different events. 

%\heng{maybe add ``potentially'', since its not proven. Could also mention future work about validating it}
The above analyses and examples suggest that PMSS can potentially be used to select the most suitable parser for downstream tasks. Further, EMSS provides fine-grained information on the parsing quality of each event cluster instead of a binary correct/incorrect. This information is especially useful for events that are less frequent or potentially indicate abnormal statuses. However, as stated in the study by Khan et al.~\cite{khan2024impact}, the number of anomaly detection datasets is limited, and the anomalies are either too obvious (e.g., Hadoop) or too latent to be discovered (e.g., OpenStack). Therefore, we did not include this part of the experiment in our paper. 

%% file: sections/8_related_works.tex
\section{Related Works}
\label{sec:related_works}

\subsection{Log Parsing}
\label{sec:background_log_parsing}
Log parsing is an essential step in log analysis; it converts unstructured log messages into semi-structured event templates to facilitate automated log analysis and reduces the effort in manual log inspections~\cite{he2021survey}. To achieve effective and efficient log parsing, many log parsers have been established. Statistic-based log parsers leverage statistical information such as frequency~\cite{dai2020logram, vaarandi2003data, nagappan2010abstracting, vaarandi2015logcluster}, similarity~\cite{fu2009execution, hamooni2016logmine, mizutani2013incremental, shima2016length, tang2011logsig}, and other heuristic rules to group log messages and extract event templates~\cite{jiang2008abstracting, du2016spell, he2017drain,makanju2009clustering, messaoudi2018search, yu2023brain,dai2020logram,dai2023pilar,liu2024xdrain,fu2022investigating,qin2025preprocessing}. On the other hand, semantic-based log parsers leverage language models for parsing. Supervised semantic-based parsers such as UniParser~\cite{liu2022uniparser}, LogPPT~\cite{le2023log}, few-shot LILAC~\cite{jiang2024lilac}, and UNLEASH~\cite{le2025unleashing} extract templates based on the knowledge of sampled labeled event templates and are highly effective. The unsupervised ones, including zero-shot LILAC, LibreLog~\cite{ma2024librelog}, LogBatcher~\cite{xiao2024free}, and LUNAR~\cite{huang2025no}, eliminated the requirement for labels and are still able to achieve high effectiveness. 

According to a large-scale evaluation conducted by Jiang et al.~\cite{jiang2024large}, statistic-based parsers are generally more efficient than semantic-based ones, but less effective in terms of event template correctness (i.e., PA and FTA). Conversely, semantic-based parsers are usually highly effective but are more resource-intensive. Further, using closed-source LLMs (e.g., ChatGPT) risks leading to privacy leaks~\cite {aghili2025protecting,sallou2024breaking}. Hence, parser selections should be based on specific circumstances. 

\subsection{Log Parser Evaluation Metrics}
\label{sec:background_statistic_based}
Parser evaluation aims to rank and select the best-performing log parser to process a system's logs. The four commonly used evaluation metrics in log parsing studies are grouping accuracy (GA)~\cite{zhu2019tools}, parsing accuracy (PA)~\cite{dai2020logram}, F1-score of grouping accuracy (FGA)~\cite{jiang2024large} and F1-score of template accuracy (FTA)~\cite{khan2022guidelines}. Using labeled ground-truths, these label-based metrics evaluate the parser's grouping and template correctness on the message and template levels. However, as discussed in Sec.~\ref{sec:challenges}, the use of these label-based evaluation metrics encounters challenges in label availability, label quality, and outcome consistency, which motivate us to seek a label-free evaluation metric. Upon analyzing the fundamental requirement of log parsing, we propose a novel metric that jointly evaluates parser grouping and template abstraction performance with the cohesion and separation of inferred event groups. 

%. First, given that log message labeling is a labor-intensive process, their reliance on ground-truth limits academic log parser evaluation to a fixed set of pre-labeled datasets. It also hinders practitioners in log parser selection for their property system, where no label is available. Second, researchers have not yet reached an agreement on label (i.e., event template) correctness~\cite{jiang2024large,khan2022guidelines,xiao2024free,huang2025no}, and the correctness cannot be verified due to the unattainability of original logging codes. Third, due to the implementation of divergent template correction rules, we found that the conclusions on parser effectiveness are inconsistent when the evaluations are conducted with different ground-truth versions. 

%% file: sections/9_validity_threat.tex
\section{Threats to Validity}
\label{sec:validity_threat}

\noindent\textbf{Internal Validity. }
We investigated 4 state-of-the-art unsupervised LLM-based parsers in our experiments. Considering LLM's stochastic nature, we collected their parsing results in one run and computed the metric scores consistently on the same output. By fixing the parser outputs and varying only the ground-truth versions, we isolate the effect of the ground-truth version on the metrics. Stemming from the same reason, supervised LLM-based parsers are excluded to avoid the impact of different label samples used in few-shot prompting.

We acknowledge that rank correlations per dataset are also critical in discussing the similarity among metrics. However, statistical tests on small sample sizes are prone to noise and can lead to biased conclusions. As an alternative, we conducted Spearman’s $\rho$ analysis on all sample points to understand the correlation between PMSS and FGA-FTA. 

%In RQ1, we aim to discover the impact of different ground-truth versions on log parser evaluation. Their discrepancy is evaluated from two aspects: absolute score differences and the change in parser with the highest scores. The first evaluation revealed that different versions can lead to large score offsets, and the second evaluation revealed that the conclusion on ``best-performing log parser'' may change when a different ground-truth version is used. The combination of the two evaluations jointly showed the challenges in supervised log parser evaluation metrics. 

\noindent\textbf{External Validity. }
To ensure generalizability, the experiments and discussions are based on 6 state-of-the-art log parsers and 14 varied-sized log files in Loghub 2.0, which were collected from different systems (e.g., distributed systems and supercomputers). Nevertheless, further evaluations on other datasets with more complex settings (e.g., messages matching the same event having different variable lengths) are required to demonstrate the generalizability of our findings. 

In Sec.~\ref{sec:parser_selection}, we theoretically discussed how to select a parser for downstream tasks using PMSS and EMSS based on existing literature. However, due to the limitations of open-source datasets, we did not conduct experiments for downstream tasks in this work. We encourage future researchers to investigate their usage in parser selection in real-life scenarios. 

\noindent\textbf{Construct Validity. }
We did not run LibreLog in our experiments due to hardware limitations. Its parsing results on the same dataset (i.e., Loghub 2.0) are obtained from the original work. %, and the template regexes are converted to Loghub 2.0 style templates for a unified evaluation. 

%% file: sections/10_conclusion.tex
\section{Conclusion and Future Works}
\label{sec:conclusion}

Our work proposes a novel label-free template-level log parsing evaluation metric, PMSS. PMSS addresses the label availability and alignment challenges in traditional label-based evaluation. The single metric assesses both grouping and template quality, the two main requirements for log parsing. We conducted experiments on 14 widely used datasets and compared PMSS's evaluation outcomes with FGA and FTA, the two mainstream template-level label-based effectiveness metrics. The results show that tools with an optimal PMSS score also tend to have (near-) optimal FGA, and the two metrics selected the same optimal parser in 7 out of the 14 studied datasets. Further, PMSS shows a significantly strong positive correlation with FGA (Spearman's $\rho$=0.648) and FTA (Spearman's $\rho$=0.587), highlighting PMSS's potential as a label-free proxy for log parsers' performance. Although PMSS generally requires more computation time due to data processing, it remains a near-linear time complexity in most cases, making it a scalable and stable evaluation alternative.  

%Based on the analyses and experiments, we extended the discussion on our design considerations and provided guidelines for practitioners in log parser selection. 
Our metric provides a novel approach to comprehending log parser performances on different granularities with fewer manual efforts. As previously discussed, PMSS still faces several challenges in data processing, and its impact on parser selection for specific downstream tasks is yet to be examined. Our future work includes a more in-depth discussion on PMSS's effectiveness (e.g., the ranking correlation with FGA and FTA per dataset), improving its data processing approaches, and investigating PMSS and EMSS’s usage in parser selection in real-life scenarios.

%% file: references.bib
@article{aghili2025protecting,
  title={Protecting Privacy in Software Logs: What Should Be Anonymized?},
  author={Aghili, Roozbeh and Li, Heng and Khomh, Foutse},
  journal={Proceedings of the ACM on Software Engineering},
  volume={2},
  number={FSE},
  pages={1317--1338},
  year={2025},
  publisher={ACM New York, NY, USA}
}

@article{chen2021experience,
  title={Experience report: Deep learning-based system log analysis for anomaly detection},
  author={Chen, Zhuangbin and Liu, Jinyang and Gu, Wenwei and Su, Yuxin and Lyu, Michael R},
  journal={arXiv preprint arXiv:2107.05908},
  year={2021}
}

@inproceedings{du2017deeplog,
  title={Deeplog: Anomaly detection and diagnosis from system logs through deep learning},
  author={Du, Min and Li, Feifei and Zheng, Guineng and Srikumar, Vivek},
  booktitle={Proceedings of the 2017 ACM SIGSAC conference on computer and communications security},
  pages={1285--1298},
  year={2017}
}

@inproceedings{nagaraj2012structured,
  title={Structured comparative analysis of systems logs to diagnose performance problems},
  author={Nagaraj, Karthik and Killian, Charles and Neville, Jennifer},
  booktitle={9th USENIX Symposium on Networked Systems Design and Implementation (NSDI 12)},
  pages={353--366},
  year={2012}
}

@inproceedings{chow2014mystery,
  title={The mystery machine: End-to-end performance analysis of large-scale internet services},
  author={Chow, Michael and Meisner, David and Flinn, Jason and Peek, Daniel and Wenisch, Thomas F},
  booktitle={11th USENIX Symposium on Operating Systems Design and Implementation (OSDI 14)},
  pages={217--231},
  year={2014}
}

@inproceedings{fu2009execution,
  title={Execution anomaly detection in distributed systems through unstructured log analysis},
  author={Fu, Qiang and Lou, Jian-Guang and Wang, Yi and Li, Jiang},
  booktitle={2009 ninth IEEE international conference on data mining},
  pages={149--158},
  year={2009},
  organization={IEEE}
}

@inproceedings{he2016experience,
  title={Experience report: System log analysis for anomaly detection},
  author={He, Shilin and Zhu, Jieming and He, Pinjia and Lyu, Michael R},
  booktitle={2016 IEEE 27th international symposium on software reliability engineering (ISSRE)},
  pages={207--218},
  year={2016},
  organization={IEEE}
}

@inproceedings{fu2013contextual,
  title={Contextual analysis of program logs for understanding system behaviors},
  author={Fu, Qiang and Lou, Jian-Guang and Lin, Qingwei and Ding, Rui and Zhang, Dongmei and Xie, Tao},
  booktitle={2013 10th Working Conference on Mining Software Repositories (MSR)},
  pages={397--400},
  year={2013},
  organization={IEEE}
}

@inproceedings{yuan2012characterizing,
  title={Characterizing logging practices in open-source software},
  author={Yuan, Ding and Park, Soyeon and Zhou, Yuanyuan},
  booktitle={2012 34th international conference on software engineering (ICSE)},
  pages={102--112},
  year={2012},
  organization={IEEE}
}

@inproceedings{kuttal2011history,
  title={History repeats itself more easily when you log it: Versioning for mashups},
  author={Kuttal, Sandeep Kaur and Sarma, Anita and Rothermel, Gregg},
  booktitle={2011 IEEE symposium on visual languages and human-centric computing (VL/HCC)},
  pages={69--72},
  year={2011},
  organization={IEEE}
}

@inproceedings{qin2025preprocessing,
  title={Preprocessing is All You Need: Boosting the Performance of Log Parsers With a General Preprocessing Framework},
  author={Qin, Qiaolin and Agili, Roozbeh and Li, Heng and Merlo, Ettore},
  booktitle={2025 IEEE International Conference on Software Analysis, Evolution, and Reengineering (SANER)},
  year={2025},
  organization={IEEE}
}

@inproceedings{gadler2017mining,
  title={Mining logs to model the use of a system},
  author={Gadler, Daniele and Mairegger, Michael and Janes, Andrea and Russo, Barbara},
  booktitle={2017 ACM/IEEE International Symposium on Empirical Software Engineering and Measurement (ESEM)},
  pages={334--343},
  year={2017},
  organization={IEEE}
}

@inproceedings{he2017drain,
  title={Drain: An online log parsing approach with fixed depth tree},
  author={He, Pinjia and Zhu, Jieming and Zheng, Zibin and Lyu, Michael R},
  booktitle={2017 IEEE international conference on web services (ICWS)},
  pages={33--40},
  year={2017},
  organization={IEEE}
}

@article{dai2020logram,
  title={Logram: Efficient log parsing using $ n $ n-gram dictionaries},
  author={Dai, Hetong and Li, Heng and Chen, Che-Shao and Shang, Weiyi and Chen, Tse-Hsun},
  journal={IEEE Transactions on Software Engineering},
  volume={48},
  number={3},
  pages={879--892},
  year={2020},
  publisher={IEEE}
}

@inproceedings{nagappan2010abstracting,
  title={Abstracting log lines to log event types for mining software system logs},
  author={Nagappan, Meiyappan and Vouk, Mladen A},
  booktitle={2010 7th IEEE Working Conference on Mining Software Repositories (MSR 2010)},
  pages={114--117},
  year={2010},
  organization={IEEE}
}

@inproceedings{vaarandi2003data,
  title={A data clustering algorithm for mining patterns from event logs},
  author={Vaarandi, Risto},
  booktitle={Proceedings of the 3rd IEEE Workshop on IP Operations \& Management (IPOM 2003)(IEEE Cat. No. 03EX764)},
  pages={119--126},
  year={2003},
  organization={Ieee}
}

@inproceedings{vaarandi2015logcluster,
  title={Logcluster-a data clustering and pattern mining algorithm for event logs},
  author={Vaarandi, Risto and Pihelgas, Mauno},
  booktitle={2015 11th International conference on network and service management (CNSM)},
  pages={1--7},
  year={2015},
  organization={IEEE}
}

@inproceedings{hamooni2016logmine,
  title={Logmine: Fast pattern recognition for log analytics},
  author={Hamooni, Hossein and Debnath, Biplob and Xu, Jianwu and Zhang, Hui and Jiang, Guofei and Mueen, Abdullah},
  booktitle={Proceedings of the 25th ACM international on conference on information and knowledge management},
  pages={1573--1582},
  year={2016}
}

@inproceedings{mizutani2013incremental,
  title={Incremental mining of system log format},
  author={Mizutani, Masayoshi},
  booktitle={2013 IEEE International Conference on Services Computing},
  pages={595--602},
  year={2013},
  organization={IEEE}
}

@article{shima2016length,
  title={Length matters: Clustering system log messages using length of words},
  author={Shima, Keiichi},
  journal={arXiv preprint arXiv:1611.03213},
  year={2016}
}

@inproceedings{tang2011logsig,
  title={LogSig: Generating system events from raw textual logs},
  author={Tang, Liang and Li, Tao and Perng, Chang-Shing},
  booktitle={Proceedings of the 20th ACM international conference on Information and knowledge management},
  pages={785--794},
  year={2011}
}

@inproceedings{le2023log,
  title={Log parsing with prompt-based few-shot learning},
  author={Le, Van-Hoang and Zhang, Hongyu},
  booktitle={2023 IEEE/ACM 45th International Conference on Software Engineering (ICSE)},
  pages={2438--2449},
  year={2023},
  organization={IEEE}
}

@inproceedings{liu2022uniparser,
  title={Uniparser: A unified log parser for heterogeneous log data},
  author={Liu, Yudong and Zhang, Xu and He, Shilin and Zhang, Hongyu and Li, Liqun and Kang, Yu and Xu, Yong and Ma, Minghua and Lin, Qingwei and Dang, Yingnong and others},
  booktitle={Proceedings of the ACM Web Conference 2022},
  pages={1893--1901},
  year={2022}
}

@inproceedings{du2016spell,
  title={Spell: Streaming parsing of system event logs},
  author={Du, Min and Li, Feifei},
  booktitle={2016 IEEE 16th International Conference on Data Mining (ICDM)},
  pages={859--864},
  year={2016},
  organization={IEEE}
}

@inproceedings{makanju2009clustering,
  title={Clustering event logs using iterative partitioning},
  author={Makanju, Adetokunbo AO and Zincir-Heywood, A Nur and Milios, Evangelos E},
  booktitle={Proceedings of the 15th ACM SIGKDD international conference on Knowledge discovery and data mining},
  pages={1255--1264},
  year={2009}
}

@inproceedings{messaoudi2018search,
  title={A search-based approach for accurate identification of log message formats},
  author={Messaoudi, Salma and Panichella, Annibale and Bianculli, Domenico and Briand, Lionel and Sasnauskas, Raimondas},
  booktitle={Proceedings of the 26th Conference on Program Comprehension},
  pages={167--177},
  year={2018}
}

@article{yu2023brain,
  title={Brain: Log parsing with bidirectional parallel tree},
  author={Yu, Siyu and He, Pinjia and Chen, Ningjiang and Wu, Yifan},
  journal={IEEE Transactions on Services Computing},
  volume={16},
  number={5},
  pages={3224--3237},
  year={2023},
  publisher={IEEE}
}

@inproceedings{jiang2024large,
  title={A large-scale evaluation for log parsing techniques: How far are we?},
  author={Jiang, Zhihan and Liu, Jinyang and Huang, Junjie and Li, Yichen and Huo, Yintong and Gu, Jiazhen and Chen, Zhuangbin and Zhu, Jieming and Lyu, Michael R},
  booktitle={Proceedings of the 33rd ACM SIGSOFT International Symposium on Software Testing and Analysis},
  pages={223--234},
  year={2024}
}

@inproceedings{khan2022guidelines,
  title={Guidelines for assessing the accuracy of log message template identification techniques},
  author={Khan, Zanis Ali and Shin, Donghwan and Bianculli, Domenico and Briand, Lionel},
  booktitle={Proceedings of the 44th International Conference on Software Engineering},
  pages={1095--1106},
  year={2022}
}

@inproceedings{zhu2019tools,
  title={Tools and benchmarks for automated log parsing},
  author={Zhu, Jieming and He, Shilin and Liu, Jinyang and He, Pinjia and Xie, Qi and Zheng, Zibin and Lyu, Michael R},
  booktitle={2019 IEEE/ACM 41st International Conference on Software Engineering: Software Engineering in Practice (ICSE-SEIP)},
  pages={121--130},
  year={2019},
  organization={IEEE}
}

@inproceedings{jiang2008abstracting,
  title={Abstracting execution logs to execution events for enterprise applications (short paper)},
  author={Jiang, Zhen Ming and Hassan, Ahmed E and Flora, Parminder and Hamann, Gilbert},
  booktitle={2008 The Eighth International Conference on Quality Software},
  pages={181--186},
  year={2008},
  organization={IEEE}
}

@inproceedings{zhu2023loghub,
  title={Loghub: A large collection of system log datasets for ai-driven log analytics},
  author={Zhu, Jieming and He, Shilin and He, Pinjia and Liu, Jinyang and Lyu, Michael R},
  booktitle={2023 IEEE 34th International Symposium on Software Reliability Engineering (ISSRE)},
  pages={355--366},
  year={2023},
  organization={IEEE}
}

@article{jiang2024lilac,
  title={Lilac: Log parsing using llms with adaptive parsing cache},
  author={Jiang, Zhihan and Liu, Jinyang and Chen, Zhuangbin and Li, Yichen and Huang, Junjie and Huo, Yintong and He, Pinjia and Gu, Jiazhen and Lyu, Michael R},
  journal={Proceedings of the ACM on Software Engineering},
  volume={1},
  number={FSE},
  pages={137--160},
  year={2024},
  publisher={ACM New York, NY, USA}
}

@article{ma2024librelog,
  title={LibreLog: Accurate and Efficient Unsupervised Log Parsing Using Open-Source Large Language Models},
  author={Ma, Zeyang and Kim, Dong Jae and Chen, Tse-Hsun},
  journal={arXiv preprint arXiv:2408.01585},
  year={2024}
}

@inproceedings{fu2022investigating,
  title={Investigating and improving log parsing in practice},
  author={Fu, Ying and Yan, Meng and Xu, Jian and Li, Jianguo and Liu, Zhongxin and Zhang, Xiaohong and Yang, Dan},
  booktitle={Proceedings of the 30th ACM Joint European Software Engineering Conference and Symposium on the Foundations of Software Engineering},
  pages={1566--1577},
  year={2022}
}

@inproceedings{le2025unleashing,
  title={Unleashing the True Potential of Semantic-based Log Parsing with Pre-trained Language Models},
  author={Le, Van-Hoang and Zhang, Hongyu and Xiao, Yi},
  booktitle={2025 IEEE/ACM 47th International Conference on Software Engineering (ICSE)},
  pages={711--711},
  year={2025},
  organization={IEEE Computer Society}
}

@article{huang2025no,
  title={No More Labelled Examples? An Unsupervised Log Parser with LLMs},
  author={Huang, Junjie and Jiang, Zhihan and Chen, Zhuangbin and Lyu, Michael},
  journal={Proceedings of the ACM on Software Engineering},
  volume={2},
  number={FSE},
  pages={2406--2429},
  year={2025},
  publisher={ACM New York, NY, USA}
}

@inproceedings{xiao2024free,
  title={free: Towards more practical log parsing with large language models},
  author={Xiao, Yi and Le, Van-Hoang and Zhang, Hongyu},
  booktitle={Proceedings of the 39th IEEE/ACM International Conference on Automated Software Engineering},
  pages={153--165},
  year={2024}
}

@inproceedings{sallou2024breaking,
  title={Breaking the silence: the threats of using llms in software engineering},
  author={Sallou, June and Durieux, Thomas and Panichella, Annibale},
  booktitle={Proceedings of the 2024 ACM/IEEE 44th International Conference on Software Engineering: New Ideas and Emerging Results},
  pages={102--106},
  year={2024}
}

@article{liu2024xdrain,
  title={XDrain: Effective log parsing in log streams using fixed-depth forest},
  author={Liu, Changjian and Tian, Yang and Yu, Siyu and Gao, Donghui and Wu, Yifan and Huang, Suqun and Hu, Xiaochun and Chen, Ningjiang},
  journal={Information and Software Technology},
  volume={176},
  pages={107546},
  year={2024},
  publisher={Elsevier}
}

@article{van2003new,
  title={A new partitioning around medoids algorithm},
  author={Van der Laan, Mark and Pollard, Katherine and Bryan, Jennifer},
  journal={Journal of Statistical Computation and Simulation},
  volume={73},
  number={8},
  pages={575--584},
  year={2003},
  publisher={Taylor \& Francis}
}

@inproceedings{lcvenshtcin1966binary,
  title={Binary coors capable or ‘correcting deletions, insertions, and reversals},
  author={Lcvenshtcin, VI},
  booktitle={Soviet physics-doklady},
  volume={10},
  number={8},
  year={1966}
}

@article{wang2020measurement,
  title={Measurement of text similarity: a survey},
  author={Wang, Jiapeng and Dong, Yihong},
  journal={Information},
  volume={11},
  number={9},
  pages={421},
  year={2020},
  publisher={MDPI}
}

@article{khan2024impact,
  title={Impact of log parsing on deep learning-based anomaly detection},
  author={Khan, Zanis Ali and Shin, Donghwan and Bianculli, Domenico and Briand, Lionel C},
  journal={Empirical Software Engineering},
  volume={29},
  number={6},
  pages={139},
  year={2024},
  publisher={Springer}
}

@misc{PMSS,
  title = {Label Free Metric for Log Parser Evaluation},
  author = {{MOOSE Lab}},
  year = {2025},
  note = {Last accessed 10 December 2025},
  url = {https://github.com/mooselab/Label-Free-Metric-for-Log-Parser-Evaluation}
}

@misc{loghub2.0,
  title = {Loghub-2.0: a collection of large-scale datasets for log parsing},
  author = {{LogPAI}},
  year = {2023},
  note = {Last accessed 10 December 2025},
  url = {https://zenodo.org/records/8275861}
}

@inproceedings{dai2023pilar,
  title={PILAR: Studying and mitigating the influence of configurations on log parsing},
  author={Dai, Hetong and Tang, Yiming and Li, Heng and Shang, Weiyi},
  booktitle={2023 IEEE/ACM 45th International Conference on Software Engineering (ICSE)},
  pages={818--829},
  year={2023},
  organization={IEEE}
}

@inproceedings{lu2017log,
  title={Log-based abnormal task detection and root cause analysis for spark},
  author={Lu, Siyang and Rao, BingBing and Wei, Xiang and Tak, Byungchul and Wang, Long and Wang, Liqiang},
  booktitle={2017 IEEE International Conference on Web Services (ICWS)},
  pages={389--396},
  year={2017},
  organization={IEEE}
}

@article{wu2023effectiveness,
  title={On the effectiveness of log representation for log-based anomaly detection},
  author={Wu, Xingfang and Li, Heng and Khomh, Foutse},
  journal={Empirical Software Engineering},
  volume={28},
  number={6},
  pages={137},
  year={2023},
  publisher={Springer}
}

@article{liao2020using,
  title={Using black-box performance models to detect performance regressions under varying workloads: an empirical study},
  author={Liao, Lizhi and Chen, Jinfu and Li, Heng and Zeng, Yi and Shang, Weiyi and Guo, Jianmei and Sporea, Catalin and Toma, Andrei and Sajedi, Sarah},
  journal={Empirical Software Engineering},
  volume={25},
  pages={4130--4160},
  year={2020},
  publisher={Springer}
}

@article{he2021survey,
  title={A survey on automated log analysis for reliability engineering},
  author={He, Shilin and He, Pinjia and Chen, Zhuangbin and Yang, Tianyi and Su, Yuxin and Lyu, Michael R},
  journal={ACM computing surveys (CSUR)},
  volume={54},
  number={6},
  pages={1--37},
  year={2021},
  publisher={ACM New York, NY, USA}
}
